\documentclass{article}

\usepackage{graphicx}
\graphicspath{ {./images/} }
\usepackage[utf8]{inputenc} 
\usepackage[T1]{fontenc}    
\usepackage{hyperref}       
\usepackage{url}            
\usepackage{booktabs}       
\usepackage{amsfonts}       
\usepackage{nicefrac}       
\usepackage{microtype}      
\usepackage{lipsum}

\usepackage{authblk}

\title{Tracking on the Web, Mobile and the Internet-of-Things}

\author{Reuben Binns\thanks{reuben.binns@cs.ox.ac.uk}}
\affil{University of Oxford, UK} 

\begin{document}
\maketitle

\begin{abstract}
`Tracking' is the collection of data about an individual's activity across multiple distinct contexts and the retention, use, or sharing of data derived from that activity outside the context in which it occurred. This paper aims to introduce tracking on the web, smartphones, and the Internet of Things, to an audience with little or no previous knowledge. It covers these topics primarily from the perspective of computer science and human-computer interaction, but also includes relevant law and policy aspects. Rather than a systematic literature review, it aims to provide an over-arching narrative spanning this large research space.

Section 1 introduces the concept of tracking. Section 2 provides a short history of the major developments of tracking on the web. Section 3 presents research covering the detection, measurement and analysis of web tracking technologies. Section 4 delves into the countermeasures against web tracking and mechanisms that have been proposed to allow users to control and limit tracking, as well as studies into end-user perspectives on tracking. Section 5 focuses on tracking on `smart' devices including smartphones and the internet of things. Section 6 covers emerging issues affecting the future of tracking across these different platforms.
\end{abstract}

\tableofcontents

\newpage

\section{Introduction}

A working definition of tracking which aligns with the focus of this paper was provided by the World Wide Web Consortium (W3C)'s Tracking Protection Working Group in 2019 \footnote{\url{https://perma.cc/3HXA-F47U}}:

\begin{quote}
    Tracking is the collection of data regarding a particular user's activity across multiple distinct contexts and the retention, use, or sharing of data derived from that activity outside the context in which it occurred. A context is a set of resources that are controlled by the same party or jointly controlled by a set of parties.
\end{quote}

While this definition is proposed within the context of the web, it could meaningfully be applied to other technologies and platforms such as mobile apps and other devices connected to the internet (the so-called Internet-of-Things or IoT). As we shall see later, similar issues arise across all three.

Let us begin by unpacking this definition with an example. Imagine Alice visits the website of a book shop and browses through their collection of wildlife books. She then goes to her favourite search engine and to search for articles about climate change policy. She sees a link to a speech made in Parliament referring to the climate crisis, which she clicks on and reads. Data about Alice's activity --- browsing the book site, entering search terms, slowly scrolling through the parliamentary records --- can be and almost certainly is being collected in some form by the organisations behind these websites and services. Such collection would not, by itself, be considered `tracking', so long as data collected within one context stayed within that context. But if an interested party somehow collates these different data points --- e.g. her book shop browsing is somehow connected to data about her search terms, or what she was looking at on the parliament website --- then this would count as tracking according to the above definition.

There are many different ways this tracking could be happening; many different parties that might be involved; and many different purposes for doing so. Alice might be tracked by her own browser, which monitors her browsing behaviour to personalise web content recommendations to her on the web and target her on other platforms (e.g. sponsored posts on a social network). She might be tracked by the search engine, which builds a picture of which search results Alice actually clicks on, so that the kinds of sites she visits show up higher in her personalised search results next time. Or she might be tracked by an advertising technology (adtech) company, whose tracking capabilities are bundled up in the code embedded by websites to display adverts and generate revenue. This enables the adtech company, who Alice has probably never heard of, to target ads to Alice based on her past behaviour on multiple different websites. Some of these vectors for tracking --- the browser, the search engine, the adtech company --- might also be owned and operated by the same company, enabling it to track Alice's activities in multiple ways.

With the advent of smartphones and internet-of-things devices, the vectors for tracking have increased; now Alice might be tracked in physical space by the apps (and operating system) accessing the GPS system on her phone, and her conversations might be listened in on by the smart speaker in her living room. All of this activity is increasingly tied together across these different devices to build ever-more detailed and proliferating personal profiles.

The term `tracking' is used to mean different things in a range of contexts, including state surveillance, public health, policing and elsewhere. This paper primarily focuses on tracking as a near-ubiquitous commercial practice which emerged through a symbiotic (or arguably, parasitic) relationship with websites, mobile apps and other internet-based services. The tracking infrastructure embedded in modern devices provides deep, intimate portraits of our lives which is already routinely used to persuade and discriminate between consumers\cite{christl2017companies}. It goes further than the most intrusive forms of government surveillance that existed before it, relying not on manual, human listening but rather on automated data capture or what Roger Clarke calls `dataveillance'\cite{clarke1988information}. Tracking could well be considered the workhorse of what some have called `surveillance capitalism'; without it, the vast wealth and and power of large digital platforms would not have been possible \cite{foster2014surveillance,zuboff2015big}. This article focuses primarily on tracking in the European and north American context, but tracking has developed differently in different parts of the world\cite{fruchter2015variations,o2021four}, and is more or less integrated within systems of state control and surveillance under different regimes \cite{yu-2009-media, stockmann-2013-media, zhou-2006-historicizing, yang-2013-power}.

The remainder of this article is structured as follows. Section 1 introduces the concept of tracking. Section 2 provides a short history of the major developments of tracking on the web. Section 3 presents research covering the detection, measurement and analysis of web tracking technologies. Section 4 delves into the countermeasures against web tracking and mechanisms that have been proposed to allow users to control and limit tracking, as well as studies into end-user perspectives on tracking. Section 5 focuses on tracking on `smart' devices including smartphones and the internet of things. Section 6 covers emerging issues affecting the future of tracking across these different platforms.

\section{Tracking on the Web}

The kind of tracking covered in this article first began on the World Wide Web (henceforth: the web), and it remains one of the main platforms on which tracking occurs. This section provides a brief history and background information on the development of web tracking.

\subsection{Pre-Web Tracking}

Early computers had little capability or reason to track users, because users were generally also programmers, and their input was limited to running programs from start to finish, as they could not be stored in memory. As systems advanced and were capable of responding to dynamic user inputs, it became useful to modify how software behaves in response to what the user had done in the past. In particular, early work on natural language dialog systems provided a variety of methods for personalising interactions with users. `User modelling' refers to the practice of generating a generic data schema in which particular user profiles can be represented, or `the construction and use of an explicit model of the user's beliefs, goals and plans' \cite{wahlster1989user}.

In some cases, this involved inferring what knowledge has already been imparted to the user (e.g. KNOME: Modeling What the User Knows \cite{chin1989knome}), while in others, it involved the construction of `stereotypes' based on analysing which user properties most often co-occur \cite{chin1989knome}. Other approaches allowed more dynamic user profiles to be built over time in response to user behaviours \cite{mctear1993user}. But these examples of user profiling did not involve user \emph{tracking} as such, because the profile was only associated with the program in question. As such, data was only collected within a single context, rather than combined across multiple contexts, and typically directly served the user's interests and needs regarding the operation of the software.

This began to change with the rise of the web. The web was one of several hypertext systems at the time\cite{berners1999weaving}, but rapidly grew to become the primary means by which the vast majority of people accessed the internet for the decades to come. 
Early website providers were in a very different position to previous software providers. Users were no longer interacting with individual pieces of software, but accessing many different websites over the internet via a single web browser. The web presented new challenges for user modelling. Since the early web was designed to exchange static documents, rather than dynamic code, there was no way for a website to record anything about a user - the website was retrieved by the user, but information about the user was largely not collected by the website. Furthermore, even if websites had a way to record user information, each individual website could only build a limited user profile based on their activity on that site.

\subsection{Surveillance-Based Advertising}
As the web developed, technology was invented to support tracking. Web developers wanted a way to be able to remember users. In 1994 John Montully, a developer at Netscape, an early web browser, created the `cookie'; a way to store a small amount of information associated with a user \cite{schwartz2001giving}. The cookie is generated by the website and stored on the user's browser. As the user goes from one page on the website to another, the browser presents the same cookie back to the web server, which can then tie the various page visits together into a coherent user `session'. This gave websites the ability to remember their visitors, with all kinds of useful applications; showing the user what has changed since they last visited, remembering what items are in the user's shopping cart, giving the user a tailored experience, and more. 

A cookie is a file containing a series of fields, including: a `name' to identify the cookie; an expiration date after which the cookie should not be sent by the browser; a domain restricting where the cookie can be sent; a URL `path' restricting where the cookie can be sent within that domain; amongst other fields. Hosts (whether first or third-parties) send cookies in response to HTTP requests from browsers, or via Javascript code. Unless the cookie is blocked by the browser, the browser will store the cookie and add it to any requests it makes to the domain defined in in the cookie, until the cookie has expired as per its expiration date.

The original purpose of a cookie --- to record a user's state (e.g. what's in their virtual shopping basket, or whether they have visited a linked page already) --- doesn't necessarily require the user to be identified. A cookie could just describe something about the user's state in a simple way. Indeed, cookies were initially envisioned as simple key-value pairs to record basic state information. For instance, a cookie designed to enable a user to resume watching a video where they left off in a previous session might have the format `resume = [time]'. But today, what is actually contained in cookies is often much more complex.\footnote{For instance, rather than generic cookies which do the same thing for different users, cookies can vary significantly between users and their purpose and behaviour cannot always be inferred from their name. This makes it harder for browsers and anti-tracking tool makers to decide which cookies to block and which to allow. A 2017 study of network traffic from real users examined the contents of cookies, finding it was often quite sophisticated and varied between users, and in many cases, the Name field differed between users, partly to act as a unique user identifier \cite{gonzalez2017cookie}.}

The news media soon picked up on this technology, with the first mention in the press apparently in 1996 \cite{jackson1996bug}. Around the same time, a new business model began to develop, on which much of the web still depends: advertising.\footnote{For details on this early history, see \cite{leiner-2009-brief, jones2020cookies}} At first, advertising on the web worked like advertising in a newspaper; the advertiser paid the website operator to display a particular advert on their page. The same advert would be placed in the same place on a web page, and seen by all visitors, for however long the advertiser paid. But unlike paper, the increasingly complex and dynamic technology of the web offered the ability to make advertising more tailored. First, rather than giving their adverts to first-party websites to distribute on their behalf directly, advertising agencies could have the first-party website include a piece of code which would retrieve for the user the latest version of the advert from the agency's own server. This way, the advertising agency had more control over their advert inventory, and could build in mechanisms to more carefully manage their spending.

To get their money's worth, advertisers would rather pay a website for each time their ad is clicked on by a user (or shown to them), rather than a flat fee for displaying the ad over a particular time period. But even better, they'd pay more for `impressions' (the term for when an ad is shown on a loaded web page) shown to users who might actually engage with their advert. If Alice has never seen an advert before, that impression may be worth more than if she's already seen the same advert 10 times before. Furthermore, if Alice is more interested in ornithology than astronomy, the opportunity to advertise to her may be more valuable to an advertiser trying to sell binoculars than to one trying to sell telescopes. To this end, advertising agencies augmented the code they gave to first-party websites for distributing banner adverts, to allow them to better track who was seeing which ads. An early adopter was the web advertising technology company DoubleClick, founded in 1995; it went on to become the biggest ad serving platform, and was acquired by Google in 2008. In setting up a direct third-party channel to observe user behaviour, companies like DoubleClick set in motion a trend that would come to dominate the web. The privacy implications of this were significant, and discussed by web developers on mailing lists as early as 1995\cite{jones2020cookies}. Further alarms were raised by privacy scholars and activists \cite{felten-2000-timing}, but their warnings largely went unheeded over the following decade.\footnote{As explained by Arvind Narayanan: ``Privacy scholars and activists were worried about surveillance capitalism at least as far back as 1995, when DoubleClick was founded. We tried to warn the public before it was too late, but mostly failed, at least in the US"  \url{https://perma.cc/Z7YF-EY3C }} From their inception, cookies were understood to involve tradeoffs between functionality and surveillance capability\cite{jones2020cookies}.

From the distribution of third-party advertiser code on first-party websites, it was a small step to the use of `third-party' cookies. First party cookies are created and set by the website being visited and stored in the browser under the domain associated with that website. Third party cookies work much the same as first-party cookies, but are set and read by third-parties via scripts embedded in first-party websites. They allow the third-party adtech firms not only to measure who sees their adverts on first-party sites, but also \emph{track} those users from one site to the next. The privacy threats of third-party cookies did not go unnoticed. The Internet Engineering Taskforce (IETF) initially proposed to outright prohibit the use of third-party cookies, or at least have them disabled by default.\footnote{See \url{https://perma.cc/Q3X5-J9AX}} However, they were already in significant use in the wild before the relevant standard was adopted in 1997 (RFC 2109), and subsequent standardisation efforts dropped this proposal. Cookies then came to be the primary technology for tracking users as they browsed the web.


In the 2000s, two things changed the relationship between third-parties and first-parties: the evolution of new programmatic advertising technology, and the increasing reliance on third-party services and resources by first-party websites. These developments are discussed in detail in the following sections.

\subsection{Programmatic Advertising}

Around 2001, advertising technology intermediaries saw opportunities to make the matching of buyers and sellers of ad space more dynamic. Rather than advertisers negotiating with publishers directly to buy ad space, adtech intermediaries could gather together multiple advertisers and publishers and dynamically match them. This enabled publishers to make money from `remnant' ad space which had not already been purchased directly by advertisers --- i.e. web pages with otherwise empty space which would be graced with valuable consumer eyeballs.

But it was hard for advertisers to tell exactly where their ads were appearing and how valuable the eyeballs on those remnant pages really were. By 2007, a new kind of intermediary arrived: the programmatic ad exchange, coupled with real-time bidding (RTB) and fine-grained personal information about the end user \cite{yuan2013real,yuan2014survey}. Ad exchanges run auctions for the ad space on a web page each time the page loads; and advertisers bid to buy that impression. This promised advertisers something quite unique; the ability to bid against each other for the opportunity to show a particular advert to a particular user based on a profile of that user. The advertiser's bidding strategies are all programmed ahead of time (often with the help of `demand-side' platforms), meaning that the auction can take place within the milliseconds which elapse between the user requesting the page and the page loading. Within this tiny window, a surprisingly large number of actors are involved in a complex, automatically choreographed bidding war, involving the processing of lots of data from multiple sources. 


The precise details of the bidding process differ from one case to another, however, on the web, there are now two main technical specifications for how these automated auctions take place. One is controlled by Google, called `Authorized buyers', and the other by the Interactive Advertising Bureau (IAB), called `OpenRTB / AdCom'. Both processes consists of a few broad steps:

\begin{enumerate}
    \item First, the ad exchange lists the auction, by broadcasting a `bid request', which contains information including: the URL of the page being visited; the site category; the user's device and browser details; one or more user identifiers that bidders can use to recognise a user from another context; the ad exchange's own cookie information about that user, which could include inferred gender, interests, location (city, or even GPS co-ordinates), and more.
    \item Then, the various competing advertisers' programmatic bidding strategies are automatically executed by their respective demand-side platform intermediaries. These bidding strategies are typically informed by further data about the user they have been able to glean from other tracking sources. Most advertisers use intermediaries called `demand-side platforms' to plan and execute bidding strategies on their behalf; they may also use `data management platforms' which help manage the user data they collect and associated ad targeting segments.
    \item Finally, the highest-bidding advertiser wins the auction, and has an advert automatically chosen, retrieved from their ad inventory, and loaded on the page for the user to see and perhaps click on (or perhaps, to ignore or block - see section \ref{section:countermeasures}).
\end{enumerate}

This new way of targeting ads led to a flourishing of new intermediaries for tracking, to meet the demand of advertisers to know more about the users whose attention they were paying to reach. The more data available to target the ad, the more the advertiser may be willing to pay to extract the maximum value from an advertising opportunity. It was therefore in the interests of various third-parties to gather as much data as they could about individual users, so as to better inform bidding strategies of advertisers, and for publishers to increase the value of their ad space when high-value users visit their sites. By building ever-more detailed profiles of increasing numbers of individuals, intermediaries could convince advertisers to pay more for the supposedly more accurate targeting opportunities. Of course, the advertisers also wanted to know that their money was being well spent; this led to the creation of a substantial sector of ad measurement providers, who assembled complex data supply chains dedicated to following what people do after seeing an ad, to establish how effective the targeted advertising actually was in driving sales and other behaviours.

\subsection{Third-party Services}

The other key development through the 2000s was the increasing reliance on third-party services and resources in web development. This came in part because the web was moving from a largely read-only web of \emph{documents}, to a web of interactive \emph{applications}. This meant web developers were no longer just formatting documents in HTML, they were now using increasingly complex code to design interactive applications in Javascript. Rather than creating all functionality of a website from scratch, it is far easier to take existing snippets from other websites and incorporate them into your own. In itself, such copying was not novel; from the very beginning of the web, developers have always liberally copied from each other (indeed, it is one of the factors which allowed the web to grow and attract users who also produced new content\cite{zittrain2008future}). However, traditionally, when web developers copied some code from elsewhere, they would still \emph{host} it from their own domain, so that when a user connects to \url{example.com}, all the HTML pages, images, Javascript files and so on, would be served via \url{example.com} as a first-party. As a result, whoever wrote the original code copied by \url{example.com} would not have any direct connection to the user's browser. As such, use of third-party code didn't facilitate third-party tracking.

However, web development increasingly came to rely on third-parties not just for code and content itself, but also for its distribution; this allowed first-parties to not worry about maintaining and hosting it all themselves. This included everything from widgets for embedding and serving video content, user analytics, the loading of special fonts, comment functions below news articles, buttons which allowed content to be shared via the user's account on a social network, anti-spam measures which separate bots from genuine human users, and much else besides. These services are typically made available through Application Programming Interfaces (APIs), which are intermediary pieces of software which allow two applications to work together. Web development moved away from first-parties bringing together all the resources for a website and hosting them themselves via their own domain and server, towards first-parties being merely the co-ordinators of increasingly complex tapestries of third-party code and content woven together and served from their respective third-party APIs \cite{zhou2015understanding}. Referencing externally hosted third-party resources had several advantages for first parties over hosting such resources themselves, in addition to saving them upfront effort and cost. These include latency (users can retrieve the third-party resource more quickly from an optimally located server), better caching (if many first-parties use the same third-party resource, users may have it cached in their browser already), and less work for the first-party to update and secure a variety of different elements.

However, while providing such benefits to first parties, third party resources amassed a significant amount of power to ingest data \cite{mayer-2012-third}. Every function outsourced to a third party opened up another network connection between the user and a third-party server, giving the third-parties the opportunity --- and in some cases a seemingly legitimate excuse --- to gather behavioural data about the user. With a venture-capital driven charge for the mass collection of personal data, these third-party services increasingly took up that opportunity. As a result, users were not only being tracked directly via the third-party scripts belonging to advertising technology providers, but also by other third-parties who were providing other, non-advertising services to first-party websites. But before long, those other third-parties also took up the opportunity to become enmeshed in the digital advertising industry too, for instance by operating as data management platforms, supplementing demand-side platforms with additional personal data about individuals to better inform ad targeting and bidding strategies. Third-parties shared some of the profits of this data monetisation with the first-parties on whom they depended, and soon the dependency was reversed; many websites came to rely on these third-parties not only for functionality (analytics, content distribution, interactive elements), but also as their main revenue source.

Among the third-party resources commonly referenced by first-parties are those belonging to the largest web giants, the likes of Google, Amazon, Facebook, Twitter, and Microsoft. This means that users of their services can be tracked as they move around the web, often even when not logged in to the respective service. This has further enriched the already rich profiles these platforms had built on their users activities beyond their walled gardens. As a result, a social network might be able to infer sensitive private information about a user not because of what they do \emph{on that social network}, but as a result of the websites they visit \emph{outside} of it. These trackers are sometimes called `personal trackers' because while they are `third-parties' in a technical sense, the user may also have a personal account with them because they also operate as user-facing first-party services like Google or Facebook \cite{lerner-2016-internet}. Whether the average user would \emph{expect} or \emph{consent} to be tracked beyond each respective service in this way is another matter. Even those without an account on a social network might still have their activity on the web tracked via its third-party tracking network; the term `shadow profile' refers to the profiles of individuals who have never created an account on a social network but nonetheless have their activities monitored and collated by them \cite{garcia2017leaking}. The same is true for all the other web giants with significant third-party prevalence on websites. 

Another complicating factor is that some third-party services may not initially be designed for tracking, but can over time become trackers. For instance, a third-party analytics service might initially be used in a way that only combines user data within a session, or within a single website. The analytics provider's code might set a cookie via the first-party's domain. But at some point, the analytics code might be updated to enable multi-domain tracking --- either by owner of the first-party domain linking together behaviour from multiple different websites, or by the third-party looking to build user profiles themselves. Google Analytics, the most popular web analytics service, gives website operators the option of linking together user behaviour across multiple domains they own. At any point, website operators could use this to track users across their sites, and Google Analytics itself can also do the same thing, across all the websites who embed their analytics code (estimated around 56\% of websites as of 2021\footnote{As estimated by W3techs \url{https://perma.cc/QG6H-J4FY}}). As such, web users may find that their web browsing activities are being tracked via Google Analytics and incorporated into a profile about them, potentially even if they have never had an actual Google account, and even if they don't use Chrome (Google's web browser).\footnote{In response to an ongoing anti-trust / competition investigation by the UK Competition and Markets Authority (CMA) in 2021, Google committed to restrict the use of Chrome browsing history and Analytics data to track users for targeting and measurement of ads on Google or non-Google websites (\url{https://perma.cc/Z5ZA-G6SN})}

A study of the development of third-party trackers between 1996-2016 provides a useful historical overview of how this phenomenon has evolved. The study found that the number of third-party trackers rapidly increased on websites during this period \cite{lerner-2016-internet}. In the early 2000s, no single tracker was present on more than 10\% of top sites; but by 2016, \url{google-analytics.com} was present on nearly a third of top sites. As mentioned above, it is now present on a majority of websites. By 2016, 300 out of the 500 top websites were making calls to at least four different APIs which could be used for tracking without cookies, up from less than 50 in 2005 \cite{lerner-2016-internet}. Another study of the development of third-party web tracking from 2005-2014 found a five fold increase in the number of external requests\cite{wambach-2016-retrospective}. There are some silver linings for the privacy-concerned, however, as some early forms of third-party tracking had already diminished or died out by 2016. For instance, `forced trackers', which use automatic popup windows, peaked in the early 2000s, before browsers began automatically blocking such popups by default around 2004 \cite{lerner-2016-internet}.

Such firms became a significant part of the US and other economies \cite{acquisti2016economics}. Internet-related advertising revenue was estimated at \$326 billion during 2020, up from \$294 billion in 2019.\footnote{'What does 2020 hold for ad markets?' WPP \url{https://perma.cc/6GJB-WZSK}} A large share of this revenue is accrued by third-party trackers ~\cite{montes2015value}.



\subsection{Cookies in Detail}

The following sections detail several important developments in the use of cookies, including cookie syncing, disguising and hijacking, and non-HTTP cookies. 

\subsubsection{Cookie Syncing, Disguising, and Hijacking}

While third-party cookies were not prohibited or disabled by default as the IETF had initially proposed, one control that was in place to limit their risks was the `SameOrigin' policy. This is a policy enforced by browsers, which limits access to cookies to the entity which set them. In this context, entities are distinguished by the domain that set them --- for instance, a cookie set by \url{firstparty.com} cannot be accessed by \url{thirdparty.com} and vice-versa --- but also by the protocol they use (e.g. HTTP vs HTTPS) and the network port from which the server is operating. This means that a cookie set by \url{thirdparty.com} via HTTPS on port 80, could not be read by a host from the same domain \url{thirdparty.com} operating over HTTP on port 81, and vice-versa \cite{braun2012origin}, because while the domain matches, the network protocol and port differ.

The SameOrigin policy places a barrier on third-party trackers, because in order to track a user across multiple sites, a tracker needs to be able to both set and read cookies on each of those sites, which means each of the first-parties must include code enabling the third-party to initiate a network connection. The SameOrigin policy is intended to make it impossible for third-parties to read the cookies set by first-parties or other third-parties. So third-parties should therefore only be able to track individuals within the networks of first-party sites on which they are present.

However, SameOrigin can be circumvented by multiple trackers working together, using a technique called `cookie syncing' or `cookie matching'. If tracker A is on \url{abc.com}, and tracker B is on \url{xyz.com}, they can collaborate together to track a user across both sites. The trick is that tracker A's code, which has been embedded by \url{abc.com}, forces the user's browser to make a network request to tracker B. This network request includes a query parameter at the end of the URL, which contains the ID of the cookie that tracker A set. For instance, the request might look something like \url{trackerB.com/cookieid?=12345}. Tracker A therefore causes the browser to reveal itself to tracker B, by leaking an identifier (\url{12345}) in a request to tracker B that tracker A set. Then, tracker A and tracker B can set up a back channel through which they combine what personal data they have about the individual by matching up their respective cookie identifiers for that individual. In this way, multiple trackers can work together to expand their reach and collaboratively build more detailed profiles of users.

Cookie syncing is one of the increasingly complex behaviours exhibited by trackers in recent years \cite{lerner-2016-internet}. Empirical studies of visible cookie matching techniques by 53 firms suggest that such techniques are present in over 91\% of the pages a user visits \cite{bashir2018diffusion}. An in-depth study of cookie syncing in the wild, using a year-long weblog from 850 web users, found that 97\% of them are exposed to cookie syncing, mostly within 1 week of web browsing \cite{papadopoulos2019cookie}. The user ID gets leaked on average to 3.5 different domains. They also estimate that a user is tracked by 6.75 times \emph{more} domains as a result of cookie syncing activity than they would if third-parties were unable to collude in this way. Other studies have utilised machine learning models to uncover forms of cookie synching which are not directly observable from the client side, by observing the adverts a user recieves \cite{cook2020inferring} (such studies raise interesting methodological questions which are addressed in more detail in section \ref{section:inferring}).

Another way the SameOrigin policy and other tracking protections are circumvented is by third-parties convincing first-parties to allow them to deliver cookies through the first-party's domain, through the use of domain redirection. This is often achieved through the configuration of CNAME records, which allow external access to a domain name space \cite{dimova2021cname}. This way, the third-party is disguised; from the browser's perspective it appears to be the first-party. As a result, third-parties can track users via different first-party domains, without violating SameOrigin, while being inconspicuous in the network traffic.  However, this also means the third-party may be able to access the first-party's cookies as well --- including those used for login and other functionality --- raising security risks for the first-party as well as privacy risks for the user. More broadly, disguising third-party cookies as first-party cookies means that users, browser makers, and researchers, can't as easily tell who is really `behind' a cookie just by looking at the domain that set it.

Even if first-parties don't grant third-parties access to users via domain redirection, there are other ways in which cookies can inadvertently give away sensitive information to third-parties. Until recently, many websites did not use HTTPS, the secure version of the regular network protocol for the web, for setting and reading cookies. In some cases, only some parts of a website would be loaded over HTTPS, while other parts deemed less sensitive might be loaded over HTTP (e.g. loading custom fonts). However, many websites also used HTTP for setting and reading cookies used for user personalisation. For websites using multiple cookies for different purposes and functionality, with complex inter-dependencies, and unclear or imprecise access-control mechanisms, it was easy to inadvertently expose cookies containing sensitive information over HTTP. A study in 2016 found that an attacker exploiting this could obtain a user's home and work address and web browsing history from Google; Bing and Baidu exposed the user's search history; and Yahoo even allowed an attacker to send an email from the user's account \cite{sivakorn2016cracked}. Ad networks like Doubleclick were also found to be inadvertently revealing the user's browsing history.

\subsubsection{Non-HTTP Cookies}

So far, we have primarily focused on HTTP(S) cookies. However, the basic concept of a cookie has been extended to other parts of the web. They all have the common function of setting and retrieving data about a user on their device, whether that happens over HTTP or some other protocol. These alternatives arose in part because as browsers clamped down on HTTP cookies, trackers moved their operations into less well-policed parts of the browser that still gave them the basic ability to write and read data.

Some methods relied on the fact that browsers often use `caching' --- keeping a local copy of various data from websites in order to not have to load it fresh every time --- to track users. By hiding a unique identifier within the data that the user is caching, it becomes possible for the host to reidentify the user next time they visit by querying some of the contents in the cache, a method first documented in 2003 and examined in detail in 2006 \cite{jackson-2006-protecting}.\footnote{\url{https://perma.cc/G9ZU-HGD5}} Another method, also first documented in 2006, was to use Flash, a now largely defunct video software platform, to store cookie-like objects. Regular cookies, set using HTTP, could be easily deleted in the browser. Flash cookies, by contrast, could be set by a first or third-party, on the user's Flash player software (which was for many years the default means of watching video on the web), and remained there even if the user were to switch to a different browser \cite{benninger2006ajax}.

These non-HTTP cookies were often used to defeat attempts by users to purge cookies from their browsers. If a HTTP cookie was deleted, a backup could be retrieved (either through the local caching or Flash methods described above), a practice called `respawning'. A study in 2009 found 281 Flash cookies across 54 of the top 100 websites \cite{soltani2010flash}. However, the use of Flash cookies declined following the publication of this research; a follow up research article by the same authors in 2011 found just 100 Flash cookies on 37 of the top 100 sites \cite{ayenson-2011-flash}. However, newer methods of achieving the same goal of reconstructing deleted cookies were also found, including the use of HTML5 local storage and ETags. Increasingly sophisticated techniques were developed to enable the reconstruction of cookies from any traces left behind by the original, deleted cookie. Cookies that were `virtually irrevocable persistent' were dubbed `evercookies' by a researcher who uncovered a particularly devious example \cite{kamkar2010evercookie}. The discovery of such mechanisms eventually lead to a lawsuit and \$500,000 settlement by one of the companies deploying them, KISSmetrics, in 2013\cite{davis2013kissmetrics}. However, large scale studies in the following years revealed the practice was still rampant \cite{acar-2014-web}.


\subsection{Fingerprinting}

The methods of tracking described above all involve tying together a user, based on identifiers stored on the browser, using HTTP cookies and related technologies. However, this is inherently less reliable because the user might delete the cookies stored on their device. While the various respawning methods discussed above might allow the identifier to be re-established, any form of browser-side storage would still result in a cat-and-mouse game between trackers trying to place cookies somewhere on the device and make them stick, and anti-tracking tools trying to find and remove them. From a tracker's perspective, rather than having to set and read an explicit identifier on a browser, it would be far more effective to be able to just tell one browser from another based on their inherent differences. This is where fingerprinting and other `implicit' methods of tracking come in \cite{eckersley-2010-unique}.\footnote{For an overview of browser fingerprinting methods, see \cite{laperdrix2020browser}
}. While the term fingerprinting is the preferred term of researchers, privacy advocates, policy makers and others, the fingerprinting industry prefers to use a host of alternative terms, like `Unique User Identifiers' (UUIDs), `Pseudo-identifiers', and `Cookieless signals'. Presumably, these are chosen to sound less scary, and more obfuscatory, than `fingerprinting'.

Unlike cookies and other methods which involve creating a unique identifier and storing it on the user's device so they can be identified later on, fingerprinting aims to identify a user based on their inherent features that might uniquely pick them out of the `crowd' of other web users (`probabilistic identification'). By analogy, it's like the difference between a car's number plate (an explicit, unique identifier assigned to the vehicle), versus a description of its brand, model, colour, scratches, and other distinguishing features. Even if the owner changes the number plate, one might still recognise it because of the brand, model, colour scheme, particular shape of a dent in the bonnet, hub caps, etc. Unlike cookie-based tracking, which involves interfering with the way the browser operates, and by necessity leaves a trace by placing a cookie, fingerprinting happens more opaquely in the background, on the server-side, and so is harder to detect. Other terminology often used to make the distinction between cookie-based and fingerprint-based tracking is `stateful' vs. `stateless' (e.g. \cite{englehardt-2016-million_track}). Cookies are classed as `stateful' because they record the `state' of a user (e.g. their previous activity on or off the site) and make that state available to a first or third party; by contrast, fingerprint-based tracking is `stateless' in that it approaches each user fresh each time, and attempts to use identifying features to tie their current session to a previously identified user. Such terminology can be misleading though, in the sense that \emph{both} approaches are capable of maintaining state; the difference being whether state is maintained via a cookie stored on the browser, or via a server-side database against which people are matched based on their browser fingerprints.

Fingerprinting methods all depend on the possibility that a tracker might be able to uniquely identify a browser based on what they reveal about themselves naturally. In some cases, this includes almost-unique identifiers that a browser emits for functional purposes, such as the Internet Protocol address from which the request is being made. Often, an IP address will be unique to a single user. But even without an IP address, there is other information that browsers may give away. For instance, the particular operating system version, the combination of plug-ins installed, the screen resolution, the selection of fonts that the browser has already installed, and so on. Such information is routinely shared with websites by browsers for a variety of reasons. For instance, knowing the screen resolution allows the website to adapt its layout to fit the dimensions of the user's screen. Knowing which fonts the user already has installed means the website doesn't need to waste bandwidth sending fonts to users who already have them installed. However, all those useful characteristics in combination may be enough to uniquely distinguish one browser from another. Tools like the Electronic Frontier Foundation's `Panopticlick', or `AmIUnique' allow users to see how unique their browser's fingerprint is \cite{eckersley-2010-unique}.

Even without access to all those browser characteristics, there are other ways to track people. An early theoretical fingerprinting attack from Felten et. al \cite{felten-2000-timing}, was based on measuring the time it takes for browsers to perform an operation with or without certain cached information. Since browser performance depends partly on what information it has already cached from its history, the speed of certain operations gives away clues as to what sites have previously been visited. That inferred history is privacy-compromising in itself, and might be enough to uniquely identify an individual browser. In the years since, such sophisticated attacks are unnecessary; there are far more readily available characteristics for fingerprinting, and fingerprint data is compact (around a dozen kilobytes) so can be collected within seconds\cite{andriamilanto2020guess}.

Some methods of fingerprinting make use of the `Canvas' element, which was part of the fifth generation of the web markup language HTML, released in 2014. The canvas element was designed to allow for the dynamic rendering of shapes and images. It can be exploited by fingerprinting scripts, which use the element to draw some text with a particular combination of font, size, and background colour. The text isn't actually shown to the user, but since different computers will render the text in slightly different ways --- due to different image processing engines, compression, speed, etc. --- the fingerprinter can distinguish between different browsers by observing how they perform this operation. It then creates a cryptographic hash of this information, and stores it so the same user can be identified later on using the same method \cite{acar-2014-web}. While canvas fingerprints are not always totally unique, they do enable identification in combination with other data \cite{mowery2012pixel}, with a 2015 study finding such methods enabled unique identification of 34\% of participants in a study of 1,000 web browsers \cite{fifield2015fingerprinting}. 

Unlike the `deterministic' tracking enabled by assigned identifiers, fingerprinting is \emph{probabilistic} in the sense that the tracker might not be 100\% certain that the browser they see in one context is the same as the browser they see in another context. They consider a wide variety of characteristics that a browser might reveal about itself, in combination with each other, in order to single an individual browser out from the crowd. This generally works because most browsers have a unique set of characteristics that allow them to be distinguished from others. In this sense, browsers are much like the characters in the board game \emph{Guess Who?}; many share many common features with each other, but any two browsers can almost always be distinguished from each other based on at least one feature.

How (in)distinguishable your browser is from all the others is typically measured in terms of the concept of \emph{entropy} as used in information theory\cite{gray2011entropy}. A system with high entropy is one where you would need to ask many questions in order to provide a complete description of its state; whereas a low entropy system is one whose state could be determined by asking only relatively few questions. A high level of entropy (measured in bits) means a browser is hard to distinguish from other browsers, while low entropy means it is easy to pick it out from the crowd. Each individual browser characteristic that is revealed to a tracker will detract from the browser's overall entropy, although some characteristics are more revealing than others \cite{eckersley-2010-unique}. Another concept sometimes used in this context is that of \emph{unicity}, which is defined in terms of the proportion of individuals in a set (in this case, the proportion of browsers exposed to a tracker) that are unique \cite{de2013unique,andriamilanto2020guess}.

While fingerprinting is harder to detect than cookie-based tracking, it is possible by either focusing on the presence of third-parties who are independently known to engage in it (e.g. through the fact that they publicly advertise their services), or by looking for the tell-tale signs of fingerprint-related behaviour. An example of the first, is
a major study of three known commercial fingerprinting providers in 2013. It showed that they were already using a wide variety of fingerprinting methods, allowing them to track users without needing any client-side identifiers (e.g. cookies) \cite{nikiforakis-2013-cookieless}. An example of the second, in an early large-scale study of fingerprinting, involved detecting when third-parties were trying to obtain a list of all the fonts the browser had installed in a suspicious manner strongly indicating an attempt to fingerprint \cite{Castellucia-2013-dataharvesting2}. Later versions of HTML also provided many new opportunities for fingerprinting, with APIs that allowed access to information like a device's battery status; while it might be useful to adapt a website so as to not use up too much power on a device with an already-low battery, this information can also be abused by fingerprinters \cite{olejnik2015leaking}.

A study from 2020 revealed that fingerprinting has only increased in the years after these initial large scale studies \cite{iqbal2020fingerprinting}. Browser fingerprinting is now present on more than 10\% of the top 100,000 websites, and over a quarter of the top 10,000 websites. Fingerprinting can also be used to re-establish explicit tracking methods like cookies. Similar to cookie respawning via storing identifiers in other parts of the browser, fingerprinting can be used to restore a deleted cookie. If a user deletes a cookie, trackers can use fingerprinting to pick them up again when they show up on another site, and re-link them back to the profile they had earlier associated with the individual.

\subsection{Email-based Tracking}
While email is not a web technology, predating the web by many years, in recent decades email has come to use many of the same technologies as the web, including HTML and Javascript frameworks for formatting and making email experiences more interactive. This has a lot to with the popularity of web-based email clients, but even non-web based email clients have for many years had the capacity to interpret HTML and run embedded Javascript in emails, making email almost a subset of the web. Therefore it should come as no surprise that tracking is also very common within emails. The original purpose for email tracking was so that senders could tell whether and when their emails were opened by the recipient. This was acheived by embedding a 1x1 pixel image. Upon the recipient opening the email, this would trigger a request over the network to load the image so it could be rendered. That request could be taken as a signal that the email had been opened. This behavioural datapoint is typically combined with other behavioural data and third-party cookies, to link together the user's email reading with their web browsing. A 2018 paper  \cite{englehardt-2018-email} found hundreds of different third-parties which track email recipients using such methods. When users click on links within emails, further information about them is leaked via the URL query string (using a method similar to that involved in cookie syncing).

\section{Tracking the Trackers}

Many of the studies mentioned above involve detecting trackers in the wild, at scale. There are distinct methodological challenges involved in such studies. As these methods have evolved, they have shaped the development of research into trackers. They have also helped inform the efforts of browser and anti-tracking tool developers to identify and block tracking. This section discusses the main methodological approaches to detecting trackers; for a more comprehensive overview, see \cite{englehardt2014web}.

\subsection{Network Traffic Analysis}

The typical methodology for web tracking detection involves capturing network traffic from the browser (typically, HTTP and HTTPS requests and responses), during a browser session. In some cases, the browser session is driven by a real human user, and in others it is simulated by a bot which automatically loads and interacts with a web page. Web tracking studies have grown in scale massively. Seminal studies in this space, starting in 2006, typicaly focused on the top 100 or 1,000 websites (e.g. \cite{krishnamurthy-2006-generating,soltani2010flash}). But larger scale measurements have become possible with new approaches. For simulated user studies, there are frameworks for crawling like OpenWPM and webXray which can easily scale to millions of websites \cite{englehardt-2016-million_track,libert-2015-1m-tracking} and billions of individual pages \cite{schelter-2016-ubiquity}. Other studies leverage data from hundreds of thousands of real users who have opted to take part in studies (e.g. users of the tracker-blocking plugin Ghostery \cite{karaj2018whotracks}).

In either case, this network traffic is then typically separated into first-party and third-parties based on domains. At this point, the collected traffic logs can be inspected. Specific types of personal information - e.g. email addresses, zip codes, telephone numbers, and other data types - can be searched for within the network `payloads' (the actual contents of the data packets being exchanged over the network). In the case of HTTP traffic, such payloads can be read in the clear, and in previous times this was sufficient to monitor the majority of web traffic. Until 2011, studies of web tracking were able to mostly rely on inspection of HTTP traffic \cite{krishnamurthy-2011-privacyleakage}. The fact that so many third-party trackers did not use the more secure HTTPS protocol with transport layer encryption, meant that their cookies could be observed by any passive observer with access to the network, e.g. internet service providers, mobile broadband network operators, and national intelligence agencies. A study from 2015 examined how a passive observer would be able to link browsing activity even as a user moved from one IP address to another, by inspecting HTTP cookies transmitted over the network\cite{englehardt-2015-cookies}. They estimate that such an attacker can reconstruct 62—73\% of a typical user’s browsing history in this way. 

However, HTTP has gradually been replaced by HTTPS, which encrypts every request and response between the client and server. To observe tracking behaviour over HTTPS, decryption is therefore required using a man-in-the-middle proxy. This involves dynamically generating and installing a certificate for each hostname (e.g. \url{thirdparty.com}). The browser is configured to trust the certificate, even though it is not the correct certificate for \url{thirdparty.com}, allowing the encrypted HTTPS traffic data to be decrypted. Then the traffic can be inspected for evidence of personal information and identifiers used for tracking.

Inspecting network traffic (whether HTTP or HTTPS) in this way for every request and response, provides important evidence to help determine whether tracking is taking place. However, it may be too fine-grained, potentially missing certain forms of tracking; for instance, it may not be easy to tell whether the data sent contains a fingerprint, because it is not possible to search for fingerprints in the same way one could search for an email address or similar. The payload might have another layer of encryption in addition to HTTPS, or it might be encoded in a way which evades simple detection methods. Finally, traffic inspection may not scale well, as some tracking may only be triggered in certain circumstances, or in response to specific user behaviours which might be missed during the data collection (especially if the study uses a simulated user rather than a real human participant). However, in many circumstances it is reasonable to infer tracking is taking place even without such evidence. For instance, the presence of a known third-party tracking domain within network traffic would strongly suggest some kind of tracking is present on the site. For these reasons, rather than examining each individual network request, some studies simply detect the presence of a known tracker, based on a reference to a known tracker host (i.e. a particular domain name, or IP address).

A major challenge with this host-based approach to tracking the trackers is how to compile a list of known trackers. Identifying a tracker is not as simple as looking at the domain it operates under. This might work for large, well known trackers who operate under a single domain. But there are many different trackers out there. Ideally, there would be a comprehensive list of all domains, with a definitive answer as to whether they are `trackers' or not. Researchers, browser vendors and anti-tracking tool providers have been working towards such lists. One important resource has been the Domain Name System's WHOIS register. This is not a single list, but rather a protocol which is used to create an interlinked system of different domain name registration databases. In theory, and in reality in the early days of the web, one could look up who owns any domain, and see some basic information about them. However over the last decade or so, increasing numbers of domain owners hide behind domain name registrar privacy services which hide their details \cite{clayton2014study}. Another factor is that publicly listing all domain name owner's details --- which often constitute personal data --- is not compliant with EU data protection law. In the lead up to the enforcement of the GDPR in 2018, EU data protection regulators, in their capacity under the Article 29 Working Party (now re-constituted as the European Data Protection Board), instructed ICANN (Internet Corporation for Assigned Names and Numbers, the organisation responsible for co-ordinating DNS) to make such WHOIS information private by default. While ICANN have been attempting to develop an alternative access system for years, which would protect privacy whilst also making WHOIS information available to those with a lawful basis for processing it, the implementation of such tiered access systems by individual domain name registrars has been uneven \cite{lu2021whois}. These factors make it hard to compile tracker lists based on WHOIS data.

Despite these difficulties, several projects have compiled \textit{tracking protection lists}. Such lists have been compiled by open-source anti-tracking projects, tracker-blocking tools, and researchers. Such lists include those from Ghostery\footnote{\url{https://perma.cc/YP9V-A7WR}}, Adblock\footnote{\url{https://perma.cc/G6LM-NHDG}}, Disconnect\footnote{\url{https://perma.cc/QR9F-QYHA}}, Adblock Plus's Easylist\footnote{\url{https://perma.cc/ZJ7X-PPPX}}, the WebXray Domain Owner list\footnote{\url{https://perma.cc/LXL9-SJ3G}}, the Exodus tracker databse (focused on mobile trackers),\footnote{\url{https://perma.cc/V7YN-MCEV}} and others. This enables researchers to scale detection of web tracking, by narrowing down to count only those hosts which have been identified as trackers
(see e.g. ~\cite{englehardt2016census,yu2016tracking,binns-2018-power}. Entries on such lists may be identified manually, which should result in high precision. However, these lists may not be fully representative of the trackers in existence. They may be oriented towards the particular circumstances of the organisations and user groups who compiled them (such as geography, sites of interest, etc.).\footnote{Another limitation of tracking protection lists is that they tend to be web-centric, an issue which makes them less viable for detecting tracking on other platforms as discussed below}. They may also have different definitions of tracking; sometimes they are advertising-oriented rather than tracking in general, and some lists make exemptions for trackers who sign up to codes of conduct.\footnote{See e.g. AdBlock Plus' 'acceptable ads' scheme ~\cite{meier2014erfolgreicher}} Finally, such lists may miss important trackers which use less visible tracking methods, like fingerprinting; one study of tracking `pixels' used for fingerprinting found that popular tracker lists missed between 25-30\% of fingerprinting-based trackers\cite{fouad2018missed}.

Another methodological difficulty for web tracking research is that third-party behaviours can be quite dynamic. The presence of a script from one third-party in a web page doesn't necessarily only indicate tracking by that single party. Once the script is run, it might trigger other third-party dependencies, who in turn might trigger further connections (for instance, to facilitate cookie syncing between third-party trackers). Furthermore, this behaviour might change from one visit to the next, leading to inconsistent results between measurements in studies which only visit / crawl each page once. Various researchers have modelled such dynamic behaviour. Gomer et al. construct \emph{Referrer} graphs, where nodes represent first and third-parties, and edges represent HTTP(S) references between them\cite{gomer2013network}. Another way of modelling these networks of third-parties is via `third-party trees', which measure the inclusion relationships between domains by recording the exact provenance of HTTP(S) requests\cite{urban2020beyond}. Using this approach, Urban et al. found that a single third-party can lead to subsequent requests by eight additional third-party services, and half of these additional third-parties change between repeated visits to the first-party. They also found that crawling only the landing page of a website may give an under-estimate of the third-parties that a user would be exposed to if they browsed around the site; deeper website crawls resulted in 36\% more third parties.

\subsection{Inferring Tracker Data Flows from Ads Served}
\label{section:inferring}
It is hard to know how trackers work behind the scenes, especially where they use techniques like fingerprinting or cookie syncing which aren't observable by just inspecting the cookies applied to a browser, or where the contents of cookies are unobservable. This presents a difficult problem for researchers who want to understand what data may have been collected and how it may be used. However, where those trackers are playing a part in the behavioural targeted advertising supply chain, it may be possible to infer what they are doing by observing their downstream effects, e.g. the ads that are eventually targeted to a user.

Studies have used this approach - attempting to infer what data trackers collect from what ads are eventually shown - since at least 2012
\cite{mayer2012third, bashir-2016-tracing}. This method can also be used to detect cookie syncing. If cookie syncing is happening, then one should expect to be able to indirectly observe its effects as trackers behave differently when they know additional information about a user that they couldn't have gleaned from their own tracking network. This has been exploited in one study of cookie syncing \cite{cook2020inferring}. By setting up fake personas which automatically browse the web, exposing themselves to different ad network trackers, and then observing the bids that those networks place in programmatic ad exchange auctions, the authors were able to guess which ad networks were colluding with each other behind the scenes. For instance, say tracker A observes a web user visiting an automotive dealer website, and then subsequently tracker B --- despite never having directly observed that persona interacting with automative-related content --- wants to bid more to target a car advertisement to that user. We might then reasonably infer that tracker A and B have a back-channel where they combine their respective cookie information to inform their bidding strategies. The authors developed a machine learning model to infer these relationships, finding evidence suggesting several instances of cookie syncing that would not be observable directly from the client-side.

As Bashir et al. point out, an advantage of this approach is that it is not necessary to establish the exact method by which the data used to target the ad was collected and shared\cite{bashir-2016-tracing}. Instead, it relies on the semantics of how exchanges serve ads, rather than the specific matching mechanism. Since such methods are essentially attempts at causal inference, they could also borrow from information flow tracking methods and more generally from experimental science and statistical analysis, as explained in \cite{tschantz2015methodology}. Along similar lines, Lecuyer et al. propose \emph{XRay}, a generic system for reverse-engineering which data is being used to drive profile-based targeting using `differential correlation' \cite{lecuyer-2014-xray_differential_correlation}.

\subsection{Cross-border Tracking Comparisons}

Various research has aimed to study the international distribution of trackers. While many tracking companies are based in the US, there are also many others located elsewhere. The cross-border dimensions of tracking invite some basic questions. Where in the world are trackers located? Are web users in some parts of the world tracked more than others? The geographical spread of trackers that a web user is exposed to will be highly dependent on what sites they visit, and where in the world they are located. This has interesting legal consequences. For instance, data about an individual in one jurisdiction might be protected to a greater or lesser extent than the jurisdiction in which the tracker is based. Relatedly, the extent to which third-party trackers can be used as surveillance infrastructure by passive network traffic observers also depends on where they are located within the global flow of internet traffic. For a national intelligence agency, the effectiveness of attempts to piggyback on tracking infrastructure depends in part on where in the world the wiretap is located.

According to a 2013 study of trackers on the most popular websites used around the world, the most common location for a tracker is the US \cite{Castellucia-2013-dataharvesting2}. Even in China, the majority of trackers were found to be US-based, despite China having a significant third-party tracking industry of its own. The only country with more local trackers than US-based trackers was Russia. Variations in tracking between jurisdictions was studied by Fruchter et al. \cite{fruchter2015variations}, who studied cookies and HTTP requests from browsing sessions originating in different countries, finding significant differences between them. A study by Hu et al. in 2020 found that users in China tend to be tracked by fewer trackers, less often, than users in the US \cite{hu2020multi}.  Regarding the use of third-party trackers as surveillance infrastructure by national intelligence agencies, Englehardt et al. \cite{englehardt-2015-cookies} note that given the concentration of third-party trackers in the US, the US National Security Agency (NSA) is particularly well placed to catch foreign users in a dragnet surveillance effort incorporating them.

\subsection{Measuring Legal Compliance and Regulatory Effects}

A number of studies of web tracking combine large-scale measurement of tracking, with heuristic rules to automatically detect possible unlawful behaviour by trackers. For instance, several studies implement tools to check whether a website sets non-essential cookies without asking for consent in a valid way (in violation of the EU ePrivacy Directive) \cite{trevisan-2017-uncovering,nouwens_dark_2020,utz2019informed,matte2020cookie}. Other studies look at whether first-parties disclose the existence of third-parties in their privacy policies as required by various laws (e.g. data protection and US sectoral / state laws)\cite{libert-2018-automated_policy_auditing}. Libert et al. propose and demonstrate a general purpose search engine for collecting court-admissible forensic evidence of non-compliance by trackers\cite{libertpreserving}.

Other studies attempt to measure the impact of regulation on web tracking, e.g. whether new regulations result in changes in tracker distribution and behaviour. For instance, Urban et al. study the effect of the GDPR on web tracking \cite{urban2018unwanted}. They conclude that while the structure of entities engaging in cookie syncing has not changed, the number of third-party connections appears to have shrunk by around 40\% since the GDPR came into force.\footnote{For similar work comparing pre- and post-GDPR tracking in mobile apps, see \cite{kollnig2021before}.}

\section{Tracking Countermeasures and End-User Perspectives}
\label{section:countermeasures}

This section aims to cover some of the main approaches that have been proposed (and in some cases, adopted) to limit or prevent tracking on the web, and research into end-user perspectives on tracking. Where alternative countermeasures for smartphones and IoT exist, they will be discussed in their respective sections below.




\subsection{Tools for Notice and Consent}

The original response to address unwanted tracking was to transplant the US `notice and consent' model of privacy regulation to the web \cite{rotenberg-2001-fair,barocas-2009-notice}. A user, confronted with a website that wants to facilitate tracking by third-parties, is given notice of who these third-parties are, what data they will use, for what purposes, and can then make a choice whether to use the website or not. Given the large number of websites and even larger number of third-parties, this may not be practical. A 2008 study estimated \cite{mcdonald_cost_2008} that the US national opportunity cost for reading privacy policies was \$781 billion. Moreover, even given time, privacy policies are difficult to read \cite{grossklags2007empirical,jensen2004privacy} and even contradictory\cite{ridiculousness_2019}. Finally, the end-user privacy settings that are available are often difficult to use, and their effects are difficult for users to understand and manage \cite{leon2012johnny,lin2014modeling}. As a result, numerous efforts have been made to automate the process of conveying tracking preferences.

\subsubsection{Tracking Preference Standards (P3P \& DNT)}

These included standardisation efforts from the likes of the W3C, with standards like the Platform for Privacy Preferences (P3P), and Do Not Track (DNT). These mechanisms aimed to enable users to control the data flows between their browsers and websites and third-parties. The premise was that by giving users a way to signal to websites their preferences regarding tracking, websites could respond accordingly. Early proposals included the idea of intelligent user agents --- software running in the web browser --- which could record users' preferences regarding tracking and automatically mediate or even negotiate between them and the interests of first and third-parties who want to track them \cite{cranor2000agents}. The hope was that:

\begin{quote}
`Architectures like P3P make possible machine to machine communication ... machines can bear the cost of this negotiation [and] be our agents for protecting our privacy'  \cite{lessig1999architecture}  
\end{quote}

The P3P standard was developed at the W3C in the late 1990s, and some websites began publishing machine-readable privacy policies in the P3P format around the turn of the century. Researchers and engineers at AT\&T labs developed and tested early prototype P3P agents, but their usefulness was limited by the fact that very few websites had adopted the standard \cite{cranor2002use}. Browsers attempted to encourage websites to format their privacy policies using P3P, but these efforts backfired when websites realised they could simply provide a technically valid policy which contained no actual content \cite{leon2010token}. P3P was dropped by most browsers in the mid-2000s, with Microsoft Internet Explorer and Edge maintaining support until Windows 10 in 2015.

A later effort to standardise a method of communicating user's preferences regarding tracking was the Do Not Track standard in 2011 \cite{mayer-2011-DNT}. This was much more blunt than P3P: it simply recorded whether the user did or did not want to be `tracked' and communicated this to servers with three states: \texttt{1} to indicate the user did not want to be tracked; \texttt{0} if they did want to be tracked, and \texttt{null} if the user had not yet expressed a preference either way. This was intended to be configured once per browser, for the whole web, not on a website-by-website basis. It was hoped that this lightweight technical protocol would be complemented by a legal and policy infrastructure that would both define what constitutes `tracking', and put in place governance mechanisms to ensure that trackers obeyed the signalled instructions. According to a 2012 survey of US consumers, DNT was popular \cite{hoofnagle-2012-privacy}. But it was declared a failure as early as the following year \cite{bott-2012-dnt}. Browsers eventually dropped support for DNT around 2016. The policymaking process came to a halt in large part due to disagreements between browser vendors, users, digital rights advocates, first-parties, and trackers about how to define tracking, reflecting their different interests. Indeed, the online behavioural advertising industry appears to object to the very term `tracking', as evidenced by the Interactive Advertising Board using the term in scare quotes in a comment on a campaign by privacy activists against tracking (described as `a facile and indiscriminate condemnation of “tracking”').\footnote{'Digital AdvertisingIndustry Warns Against Misguided EU Regulation - IAB Europe' https://iabeurope.eu/all-news/digital-advertising-industry-warns-against-misguided-eu-regulation/}

A fresh attempt to create a global technical standard for privacy preference setting is the Global Privacy Control (GPC).\footnote{\url{https://perma.cc/26BK-N6S8}} Like P3P and Do Not Track, this is a browser-configured global opt-out signal; proponents argue that websites are obliged to respect GPC signals under the California Consumer Protection Act (CCPA) and the GDPR. It remains to be seen whether this latest incarnation of a global privacy preference standard will be successful, but it has been backed by the California Attorney General. Some might question whether a user interaction which indicates consent to tracking (e.g. by clicking `accept' on a consent dialogue box) should take precedence over a global browser-defined setting, since it is an affirmative action; however, the fact that such dialogue boxes often use dark patterns to coerce acceptance \cite{nouwens_dark_2020,gray2021dark} might lend the global browser setting more authority. Finally, another recent proposal in a similar vein is the `Advanced Data Protection Control' standard, proposed by NOYB, a European data rights organisation; this aims to provide for more fine-grained and flexible browser consent signals particularly suited to E.U. data protection law. Whether these new initiatives can succeed where previous ones have failed remains to be seen. One thing counting in their favour this time are explicit legal provisions for their use; e.g. for those users wishing to exercise their right to object under the GDPR, Article 21(5) ensures that `the data subject may exercise his or her right to object by automated means using technical specifications' (such as browser settings).

\subsubsection{Privacy Policy Languages}

Despite the failure of previous attempts to standardise protocols for communicating privacy and tracking preferences, researchers and others have nonetheless pursued various kinds of formal languages for describing privacy and tracking-related policies. These have a variety of uses, from summarising policies on behalf of users, to analysing and auditing the stated practices of trackers.

Some projects, like TOS-DR (`Terms of Service; Didn't Read') crowdsource the annotation and highlighting of key clauses in terms of service and privacy policies\cite{tosdr-2013}. This approach has also been explored in research, where crowd-sourced classifications of policies could be used as training data to train a machine learning classifier to recognise the content of previously unclassified privacy policies, enabling them to be summarised and made more comprehensible \cite{wilson-2016-crowdsourcing,tesfay_2018}. 

The widespread adoption of standardised or semi-standardised privacy policy formats would not only be useful for end-user consent tools. It could also enable large scale automated evaluation of different websites, which could provide insights into trends and practices, automatically identify outliers, and other kinds of analysis (such possibilities are explored in other contexts where standardised privacy notices were adopted, e.g. financial institutions \cite{cranor-2016-large}). Even when privacy policies are not standardised in this way, natural language processing techniques can be used to identify disclosures of tracking practices, e.g. \cite{libert-2018-automated_policy_auditing,privonto_2017,amos2021privacy}. Various projects aim to define formal semantic languages to represent privacy-related information; for an overview, see \cite{zhao2016privacy}, and a recent example, modelled around data protection regulation, see \cite{grunewald2021tilt}.

\subsection{First-party Limitations on Tracking}

While first-parties are often to a large extent dependent on third-parties for parts of the functionality of their websites, as well as for their revenue via advertising, there are still ways in which they can limit third-party tracking without necessarily hurting their operation and bottom line. At the very least, first-parties can audit their own websites to ensure that any user data extracted by third-parties is in line with their expectations and policies. There are several methods for monitoring data flows at runtime and ensuring they are secure \cite{hedin2014jsflow}. An early overview of such techniques and a review of the enforcement of privacy and security policies by web developers is provided in \cite{bielova2013survey}. More straightforwardly, many of the functions that are currently outsourced to third-parties could potentially be hosted by first-parties themselves, to protect the privacy of users, for example using self-hosted analytics rather than third-party analytics\cite{akkus2012non,chandler2016using}.

In the last decade, numerous companies have emerged which offer tools for first-parties to audit their own sites to identify and manage the third-party resources they use.\footnote{See e.g. QuantCast, OneTrust, TrustArc, Cookiebot, and Crownpeak.} Some use these audits to automatically generate privacy policies for the first party, which describe the third party trackers. More recently, many have offered so-called `consent management platforms'; pop-up consent boxes designed to facilitate consent interactions from the user on behalf of the first party. Whether these privacy policy-generators and consent management platforms actually meet the legal standard for consent under EU data protection law is questionable \cite{nouwens_dark_2020}; it is also unclear whether they or the first party are ultimately responsible for ensuring such compliance \cite{santos2021consent}. However, given that existing web infrastructure and web development practices are so heavily tilted towards third-parties, there is a clear need for better tools to enable first-parties to limit and manage such issues. 

\subsection{Tracker Blocking and Obfuscation}

Both tools for notice and consent, and first party limitations on tracking, rely on first parties to take action against the interests of third parties. But given that first parties are often dependent on third party services for their revenue, such action is often lacking. As a result, unilateral methods of preventing tracking, which block trackers at a technical level via the browser, have instead proven popular.

Tracker blockers have been implemented both by specialist browser plugins like Ghostery, Disconnect, Adblock, Adblock Plus, Privacy Badger, and others, as well as in-built controls within the browsers themselves.\footnote{There are also other unilateral methods which prevent tracking at the local network level; see section \ref{section:iot_counter}} Users of specialist tracker-blocking software are more privacy-motivated and digitally literate than average \cite{baruh-2017-online}, although they do not always have good mental models of how these extensions actually work \cite{mathur2018characterizing}. Regardless, tracker-blockers do provide meaningful protection with very little effort on the part of users, with users reporting only occasionally encountering websites which do not work when the tracker-blocker is enabled\cite{mathur2018characterizing}. Where specialist browser plugins have ventured to offer new, more aggressive means of tracker-blocking, browser providers have largely followed; Safari and Firefox block third party cookies by default, and Chrome has announced it will follow suit in 2022 (although similar commitments in the past have not been upheld).

Dedicated tracker blocker tools can work in a variety of ways. Some block network traffic at the operating system level, so they work beyond a specific browser.\footnote{E.g. MVPS,  “Blocking  unwanted  connections  with  a  hosts  file,”  2015, \url{https://perma.cc/NM8U-T3T6}, and P. Lowe, “Yoyo hosts file,” 2015, \url{https://perma.cc/4TR6-SVX4}.} An advantage of this is that trackers can be blocked regardless which browser or other application is being used; this could, for instance, help stop the kind of email tracking discussed above if the user were to be using a native email client. The disadvantage is that they block entire domains or subdomains, rather than specific URIs: e.g. \url{thirdparty.com} would also block \url{thirdparty.com/privacypolicy}.

However most tracker blocking is browser-based. Initially, it was not tracker-blocking but ad-blocking plugins that were popular. Ad blockers like \emph{AdBlock Plus} block the loading of programmatic web ads so users don't see them on the page. This doesn't necessarily mean they prevent tracking. Adtech platforms which both distribute ads and track users to enable ad targeting, might still collect the data even if the ad blocker blocks it from showing ads. However, tracker blockers, like \emph{Ghostery}, \emph{Privacy Badger}, and \emph{Disconnect}, display trackers that are present on a page and allow users to block them.

There are different approaches taken by different tracker blockers to determine what constitutes tracking, and how they detect trackers. As explained in section 3.1 above, there are large and relatively comprehensive lists of tracking hosts which have been built by tracker-blocking projects and researchers. An 8-year study of EasyList, one of the most popular curated lists, found that there were often errors in the list (e.g. non-tracking sites which were wrongly identified as trackers, and vice-versa) which took substantial time to be corrected despite the list owners being notified \cite{alrizah2019errors}. Community-curated lists also often contain idiosyncratic rules that are ill-defined and documented by a small number of contributors, which can lead to outdated, ineffective or superfluous tracker-blocking behaviour \cite{snyder2020filters}.

In addition to community/company derived rules, there are also algorithmic approaches, which use statistical and machine learning methods to identify trackers\cite{bau2013promising,ikram2016towards}. For instance, Privacy Badger and Ghostery both rely on an algorithmic approach, which automatically classifies trackers based on detecting similar patterns of high-entropy strings being sent by multiple first-parties to a single third-party \cite{yu2016tracking}. A study of tracker blocking in 2017 found that (some of) the rule-based trackers performed better than the algorithmic ones; that trackers with a smaller footprint (i.e. present on fewer sites) were more likely to evade tracker blockers; and that all had blind spots when it came to identifying and blocking fingerprint-based tracking, and tracking by third-party content delivery networks \cite{merzdovnik2017block}.

While earlier studies suggested that the algorithmic approach to tracker identification might be less successful than manually-curated tracker lists, more recent research and real-world deployments have used machine learning to identify trackers at scale. For instance, Iqbal et al. use a graph-based machine learning approach \cite{iqbal2020adgraph}. They represent the HTML, network requests and javascript behaviour of a webpage as a graph, and use this in combination with existing manually curated lists of trackers as ground truth to train an ML model that classifies trackers. Others have explored machine vision to detect adverts as visual elements on a page and block them\cite{tramer2019adversarial}; however, this only blocks adverts, rather than blocking tracking infrastructure in general. Apple's Safari browser has begun to deploy statistical and machine learning approaches to detect trackers, an approach it calls `Intelligent Tracking Prevention' (ITP). Using statistics on resource loads, user interactions, and other data, it infers whether a host is a tracker or not. Cookies from hosts deemed to be trackers are then blocked from third-party use after one day since the user interacted with the website, and then automatically deleted entirely after 30 days. ITP has had a real impact on trackers' bottom line. The CEO of Criteo, an adtech firm heavily reliant on third-party tracking, attributed the firms loss of over 20m revenue to ITP\footnote{\url{https://perma.cc/G694-YNVQ}}.

Tracker blocking has of course not gone unnoticed by trackers themselves, who have developed a range of \emph{counter-}countermeasures to address it. Some attempt to detect when a user is using an ad or tracker blocker, and thus enable the first-party to refuse to serve the page content until the user disables their blocker \cite{iqbal2017ad}. Others attempt to circumvent blockers. For instance, a bug in the Chrome browser meant that ad blockers could not intervene on connections opened up over WebSocket (a low-overhead alternative to standard HTTP/S); a study found that a small but persistent group of trackers exploited this bug by routing their traffic through web sockets, until the bug was fixed \cite{bashir2018tracking}. Taking the arms-race to another level, Zhu et al. aim to detect attempts by trackers to detect tracker-blockers by using `differential execution analysis'\cite{zhu2018measuring}. By comparing the behaviour of a webpage when visited with and without an adblocker, the counter-countermeasures used to detect ad blockers can be isolated, and thus \emph{counter}-counter-countermeasures can be developed.

Tracker blockers have traditionally been designed to work against cookies, and as a result have struggled to grapple with cookie-less tracking techniques (i.e. fingerprinting). Even worse, the fact that a user has installed a tracker blocker plugin might even be used by a fingerprinter as an additional identifying feature, as observed by Nikiforakis et al. \cite{nikiforakis-2013-cookieless}.

In addition to tracker-blocking tools which \emph{prevent} the flow of data to trackers, some have proposed an alternative approach of introducing deliberately false data into the tracking ecosystem, a practice known as \emph{obfuscation} \cite{brunton2015obfuscation}. The \emph{AdNauseum} project aims to do this in the context of behavioural advertising \cite{howe2017engineering}. It is a browser plugin that automatically randomly clicks on ads in the background while a user is web browsing, to pollute their inferred interest profiles with things they aren't actually interested in. Another study of obfuscation, focusing on the tracker BlueKai (owned by Oracle), found that injecting just 5\% fake traffic in addition to real traffic is enough to meaningfully alter the interest profile a tracker compiles about a user \cite{degeling2018tracking}. While such deceptive practices may appear malicious, Van Kleek et al. argue that there are many potentially beneficial uses of what they call computationally mediated pro-social deception \cite{van2016computationally}; tricking the trackers might therefore be considered within the category of `benevolent deception'.

\subsection{Privacy-preserving Alternatives}
\label{section:pp_alternatives}
One of the most commonly-cited objections to tracking is that it is highly privacy-invasive. But what if the underlying ends of tracking could be effectively achieved, in less privacy-invasive ways? Several proposals aim to do this using `privacy-preserving' techniques; they enable some of the same functions --- targeting adverts, differentiating between users, enabling aggregate insights --- but without a third-party (or even first-party) `seeing' an individual's data.

Privacy-preserving ad targeting aims to enable adverts and content to be targeted at individuals based on their behaviour and interests, without such data actually being revealed to any adtech intermediaries or advertisers. For instance, Hadadi et al. propose a system wherein a user interest profile is created locally, on their own device, and a range of possible ads are selected from locally based on that profile \cite{hadadi-2011-targeted}. Davidson et al. similarly propose local models for personalisation based on user personas on the Windows mobile operating system \cite{davidson2014morepriv}. Other proposals use techniques like private information retrieval\cite{chor1995private}, a cryptographic protocol which allows items to be retrieved from a database without revealing to the database owner what those items are\cite{ullah2017enabling}. There may even be ways to achieve ad conversion measurement in ways which don't reveal any individual user data (see e.g. \cite{green2016protocol} and Apple's `Privacy Preserving Ad Click Attribution For the Web'\footnote{\url{https://perma.cc/Q33U-VRCH}}).

Whether these approaches really address the ethical concerns with tracking is questionable. If the issue with the surveillance-based web is that it enables behavioural data to be associated with an explicit, identifiable individual --- the `privacy as confidentiality' paradigm --- then these approaches represent a promising alternative. But if the issue is instead that it enables corporations, governments and other, even less accountable actors to manipulate and misinform the public, or to discriminate against groups (even if they don't know which individuals are in those groups), or more broadly to shape the information environments and choice architectures of billions of people, then arguably it doesn't matter whether they can access personal data while doing so. As such, these privacy-preserving approaches may not be the answer to the broader political concerns that tracking raises \cite{gurses2010pets,tavani2001privacy,agrawal2021exploring}.

\subsection{End-User Perceptions, Expectations, and Choices}

A motivating assumption for much of the research into the detection and mitigation of tracking is that it presents risks to those tracked. But further investigation into how the \emph{subjects} of tracking understand, perceive, feel and respond to tracking is important to help ground empirical research, the design of countermeasures, and regulatory responses. Substantial research in this vein has been undertaken within human-computer interaction (HCI) research over recent decades \cite{cranor-2000-beyond,felt2012ve,king2012come}.

Privacy concerns appear to vary internationally \cite{bellman-2004-international}, by age cohort \cite{youn-2005-teenagers}, and between iOS and Android users\cite{benenson_android_2013}. Attitudes towards data flows are also influenced by the context(s) from which the data are drawn, as well as the type of data being disclosed, perceived norms around disclosures, individuals' relationships with the data recipient(s), and the purposes of disclosure~\cite{barua2013viewing, Bilogrevic2016, Naeini2017, klasnja2009exploring,lee2016understanding, leon2013matters, lin2012expectation}.  When their expectations are violated, e.g. by data being sent to opaque third-party trackers in unanticipated ways, people feel `creeped out', and when people become accustomed to the reality of such unwanted data flows, such feelings transform into helplessness and resignation \cite{shklovski2014leakiness,ur-2012-smart}. Attitudes to tracking are typically placed within the broader theoretical frameworks of privacy. These include conceptualisations of privacy as control over information about oneself or `informational self-determination' \cite{altman1975environment,westin1968privacy}, and `contextual integrity' \cite{nissenbaum-2010-privacy}, according to which privacy norms regulate the transmission of information between various contexts, between actors, concerning certain attributes, according to certain transmission principles appropriate to those contexts ~\cite{Apthorpe2018, Martin2016,barkhuus2012mismeasurement}.

Early research into privacy attitudes suggested that people may divide into a small set of distinct privacy `types' \cite{westin-1967-privacy,kumaraguru-2005-privacy}, including `privacy fundamentalists' who avoid disclosing their information at all costs, `pragmatists' who make calculated tradeoffs between privacy and benefit, and `unconcerned' who are willing to give away their personal information for little or no benefit. One might expect these broad categories to predict people's (intended or actual) behaviours in specific situations; however, later research has found no significant correlations between the two \cite{woodruff2014would}. Furthermore, the relationship between preferences, intentions, and behaviour is complicated and fraught, especially in the context of the ubiquitous and opaque tracking described above. Some studies purport to show discrepancies between people's stated preferences and their behaviour, such that someone who claims to care about privacy might later behave in a way which suggests they don't. Various attempts have been made to explain (and explain away) these differences \cite{Acquisti2015a,Kokolakis2017,Smith2011}, with some concluding that researching privacy in experimental isolation is unhelpful~\cite{Dourish2006}, and that privacy attitudes are relatively unpredictable across different scenarios~\cite{woodruff2014would,draper2017privacy}. While people appear unwilling to pay tangible economic costs for privacy, as evidenced by the popularity of free over paid services (and experimental results \cite{beresford2012unwillingness}), it is also not clear that paid alternatives are actually more privacy-protective\cite{han_price_2020}.

While it might seem obvious that targeted advertisements are preferable to non-targeted ones (and some research supports this\cite{chowdhury2006consumer,chanchary2015user}), this needs to be weighed against the reality of the tracking infrastructure that comes along with targeted advertisement which may be less desirable. A survey commissioned by the UK data protection regulator found that 63\% of participants found digital advertising acceptable to fund free content, but after explanation of how real-time bidding works, this fell to 36\% \cite{ico2019ofcom}. This illustrates how privacy preferences, intentions and behaviours that we observe in people need to be interpreted alongside the level of awareness, understanding, competing pressures, incentives and choice architectures that people face in real-world contexts\cite{egelman2013choice}. Stated privacy attitudes and preferences may not predict behaviour because people may lack an adequate understanding and awareness of privacy risks associated with certain technology use. Various studies demonstrate how even technically savvy users may lack a complete understanding of all the ways their data is being transmitted and used, whether via the Web, through smartphone platforms and apps, or connected devices in the home~\cite{Zheng2018a,kang2015my}. Given the complex nature of online behavioural advertising, it is no surprise that users lack accurate and complete mental models of how it works\cite{yao2017folk,eslami2018communicating}.

One of the reasons people don't take further action against privacy-violating tracking is the infeasibility or unavailability of alternatives \cite{Tabassum2019}. Without awareness of privacy controls, and more importantly, belief in their efficacy, people are unlikely to take action (a conclusion also found in relation to security behaviours\cite{Wash2015,Lee2008}). However, people's level of concern about and desire to limit data flows can change once they are made apparent in a form that users can understand. Several studies develop privacy awareness tools, which visualise data flows involved in web and mobile tracking and reflect back to people how their data is used\cite{gerber2018foxit,baruh-2017-online,weinshel2019oh,van_kleek_better_2017,van_kleek_x-ray_2018,rao2015they}. Given these tools, study participants were able to articulate more specific and actionable privacy preferences ~\cite{gerber2018foxit,baruh-2017-online,weinshel2019oh}, as well as views on ethical, economic (business models), and political dimensions of the data economy~\cite{van_kleek_better_2017,van_kleek_x-ray_2018}. 

Parents and children face particular challenges to understanding and responding to tracking. Various work seeks to understand how parents and children communicate and develop privacy preferences, with important differences between teens \cite{cranor2014parents} and younger children \cite{dowthwaite2020s,zhao2019make}. Educating children about privacy risks and improving their literacy can be effective in changing their online disclosure behaviour\cite{desimpelaere2020knowledge}, although the aforementioned lack of tools to easily control tracking may ultimately hamper effective action.

\section{Tracking on `Smart' Devices}

The web is where the practices of third-party tracking first arose, but a similar model has been deployed beyond. On smartphones, there are large app ecosystems which have their own extensive array of trackers. The advent of the `smart home' and `Internet-of-Things' devices also present new prospects for ubiquitous tracking. This section addresses tracking on these various non-web platforms and devices.

\subsection{Smartphones and Apps}

While tracking on smartphone apps has much in common and shares similar infrastructure to web-based tracking, there are some important differences.

\subsubsection{Apps vs Browsers}

The main difference between smartphones and the web is that on smartphones, the web browser is just one app amongst many. Web tracking therefore co-exists alongside app tracking on smartphones. Web tracking on mobile browsers is largely the same as on other devices, but there are some differences. 
Early studies of mobile web tracking from 2013, when smartphones were just beginning to overtake their `dumb' predecessors in rich countries, confirmed that web users were being tracked similarly regardless whether they used a mobile or desktop browser. Notable differences at the time were the comparatively smaller number of mobile-specific third-party trackers, and the (then mostly theoretical) possibility of being able to access more data from the smartphone that would otherwise be unavailable (e.g. location) \cite{eubank2013shining}.

In the years since, the availability of smartphone sensor APIs including motion (via accelerometers), position (via magnetometers), and environmental features (air temperature, humidity, light, etc), have presented an additional set of vectors for tracking mobile web users. A study of mobile web tracking found that 63\% of the scripts that access sensor APIs --- while often for a legitimate purpose - also engage in fingerprinting using those sensor readings \cite{das2018web}. These add additional bits of entropy which enhance the ability of web-based tracking on mobile web browsers.

However, the bigger difference in tracking on mobile devices is the tracking involved in native mobile applications, which have access to the aforementioned sensor data and much more. These are built on top of the Android or iOS platforms and interact with profiling technologies in different ways. While mobile apps are sometimes developed using web technologies, they are more often developed in OS-specific development environments and languages (e.g. Swift for iOS and Java for Android), rather than the standards and programming languages of the web (e.g. Javascript, HTML and CSS). The main ways in which tracking on apps differs to web tracking are: the types of data available; the app distribution model; the operating system configurability; and the use of advertising identifiers rather than cookies.

The types of data available on a mobile present different opportunities for trackers. Mobile devices frequently move around in physical space, so location and sensor data as described above provide valuable data for tracking purposes - both as a means of singling out an individual user, and to infer behaviours and interests to add to a profile for targeting purposes. Other information that might be read from a phone by an app include a list of the user's contacts in their phone book, the handset model and other device information, and a list of the other apps that the user has installed. The latter may sound innocuous but the list of apps someone uses can often uniquely identify them, making it an effective fingerprint in its own right \cite{achara2015unicity}.

The mobile app distribution model is also substantially different from the web. On the web, a web application hosted anywhere can (by and large) be accessed by anyone, largely without interference from gatekeepers. But mobile apps need to be packaged up by developers, distributed somehow, and installed by users. In theory, this could be done without going through the main official app stores run by Apple and Google; for instance, the alternative F-Droid app store for Android distributes apps not approved by Google on the official Play Store,\footnote{See \url{https://perma.cc/Z6Q8-KYMT}} and users can 'side-load' app files directly. But the vast majority of apps are installed via app stores. And in most parts of the world (outside of China, where most users use multiple app stores \cite{wang2018beyond}), most smartphone users have only one app store from which to download apps: the official app store of their chosen OS. As a result, Apple and Google have the ability to act as gatekeepers~\cite{anderson2010inglorious}. Providers have the power to effectively exclude apps based on their practices. Early studies demonstrated that this gatekeeping role was not being used to prevent apps leaking personal data\cite{egele2011pios}. But iOS and Android do limit what developers can do in various ways, and have increasingly built in greater privacy controls for end users and restrictions on app developers with respect to tracking. These controls operate at both a technical level - by restricting the ways that OS-level APIs can be utilised by apps - and at an infrastructural / contractual level by deciding which apps are allowed to appear on the iOS app store and Google Play store.

Another major difference is the increased ability of app developers to utilise the functionality of the mobile operating system compared to the web. Web browsers and web standards determine how resources are rendered, what scripts are run in what order, and so on (like the SameOrigin policy). Equivalent limits are not placed on mobile apps, which have tended to have more power to control a user's device than websites. While apps are `sandboxed' from each other to prevent apps from reading / writing between each other without permission, within their own sandbox apps can generally run whatever code and allow whatever network connections the app developer builds into their app, given the permissions the developer has requested.\footnote{Although, developers can and frequently do circumvent the permission model, through side-channels (where one app obtains data that it does not have permission for) and covert channels (where an app obtains data via another app which does have permission for such data); see \cite{reardon_50_2019}}  

At the same time, users themselves have very little ability to modify their own device, including in ways which might limit tracking by apps. User access to lower-level operating system functions generally requires an additional, non-standard layer of permission. This is known as being `rooted' (Android) or `jail-broken' (iOS). The process is not straightforward, and may void the device warranty. Many apps attempt to detect whether a device is rooted, and will not run on it if so. As a result of this, the available tracker-blocking tools on smartphones are far behind the equivalent plugins available for web browsers, although some exist through their app store policies. The ability to block tracking is directly limited by OS providers themselves in some cases. \emph{TrackerControl}, an app which reveals trackers on mobile apps, illustrates this point.\footnote{\url{https://perma.cc/73TX-SKCW}} It has two versions. One version is available on Google Play, and can show the user what trackers are present on different apps. The other version has the ability to also \emph{block} trackers, but has to be loaded from the alternative Android app marketplace `F-Droid' as the Google Play store refuses to allow apps which enable such blocking.

\subsubsection{Mobile Advertising Identifiers}
\label{section:mobile_ad_id}
Perhaps the biggest difference in smartphone tracking is the use of specialised, OS-defined advertising identifiers instead of cookies. The first generation of mobile third-party trackers would typically extract all manner of identifiers that are permanently associated with the smartphone device. These included the International Mobile Equipment Identity (IMEI) number associated with the device itself, the International Mobile Subscriber Identity (IMSI) number associated with the SIM card issued by the cellular network, the phone number, the device identifier, the MAC address used for connecting to WiFi networks, and others (see \cite{razaghpanah2018apps} for how third-parties are able to access these identifiers). The use of these different identifiers is highly problematic, since they often cannot be re-set and are strongly associated with the user and their accounts (e.g. with their cellular network account). To address this problem, both iOS and Android introduced separate `advertising identifiers' (on Android, the Android Ad ID (AAID); on iOS, the Identifier for Advertisers (IDFA)) for the purposes of third-party tracking. At the same time, they have made efforts to block off access to the other identifiers both at a technical level and through educating and warning developers against such practices, and enforcing their policies through their app stores. These Ad IDs can be re-set by the user; re-setting has the intended effect of breaking a tracker's ability to link between the user's previous and future activity.

In addition, both platforms created settings where users can limit access to the Ad ID. On iPhones, the `Limit Ad Tracking' option prevented app developers from being able to request access to the user's Ad ID. This has since been replaced by a new mechanism on iOS. From iOS version 14, under the \textit{Apple Tracking Transparency} (ATT) framework, app developers are required by Apple to request explicit consent from users for tracking. iOS users are faced with a permission screen asking them to `\emph{allow [the app] to track your activity across other companies' apps and websites}'. The user can either click `\emph{Ask App Not To Track}' or `\emph{Allow}', and both options are equally prominent on the screen. When a user selects the first option, access to the IDFA commonly used by trackers is denied (initial estimates suggest between 60 and 95\% choose this option ~\cite{att_optout1,att_optout2,att_optout3}). The move proved highly controversial with third-party trackers; prominent critics included Facebook, who paid for a series of adverts in (non-targeted, paper) newspapers extolling the importance of its tracking-based targeted advertisements for small businesses through the Covid-19 pandemic. Initial reports suggest the third-party tracking industry is right to be concerned; Facebook's own figures suggest 80\% of their users will choose to not be tracked. While circumvention is possible through alternative forms of tracking which don't rely on the IDFA (e.g. fingerprinting\footnote{See e.g. \url{https://perma.cc/46QY-PJMX}}), cutting off access to the IDFA will likely significantly disrupt current tracking practices.

Google have tended to lag behind Apple on such measures, but have gradually implemented some similar options. The Google Play Store states that the Ad ID should be used for `advertising purposes', and like the previous iOS approach, Android gives users the ability to set a preference to `opt out of personalisation'. However, the function of this setting was historically somewhat ambiguous. Developers are still technically able to access the Ad ID if the user has opted out of personalisation, but are required as a matter of Google Play Store policy to not use the Ad ID for personalised advertising purposes if this opt-out had been set by the user. However, developers could \emph{still} use the Ad ID for other purposes, such as analytics, fraud detection, and potentially other kinds of tracking, provided they have a privacy policy covering such uses.

This rather vague and difficult-to-enforce policy on Android offered less reassurance than the technical measures deployed by Apple which cut off access to the IDFA entirely for users who opted out. Starting in 2021, Google announced that they will no longer be relying on developers honouring the Play Store terms, and instead be blocking access to the Ad ID at the API level.\footnote{https://support.google.com/googleplay/android-developer/answer/6048248\#zippy=\%2Ctargeting-devices-without-an-advertising-id\%2Cpersistent-identifiers-including-android-id} Developers will be required not to connect data to a previously obtained Ad ID if the user has now removed the Ad ID from their device (although it is unclear how this will be enforced). Google also announced that they will provide alternative app-set permanent identifiers for `essential' non-advertising purposes like analytics and fraud-detection to smooth the transition away from reliance on the Ad ID. New measures will be put in place to prevent remote code being loaded during run-time that could be used to circumvent Play Store policies. From April 2022, Google will require developers to provide accurate information on the personal and sensitive data they collect. It remains to be seen how much control users will have over these alternative permanent identifiers and how closely their use will be monitored and how effectively the purpose limitations will be enforced.

The ready availability of these identifiers has meant that tracking of mobile users across apps is in many ways a much simpler affair than tracking users across the web. Cookies are unnecessary, because devices have readily give up unique identifiers across different apps, with no need for a third-party to set their own unique identifier. Fingerprinting of mobile devices is also largely unnecessary for the same reason, although some trackers do now offer mobile fingerprinting services and make use of mobile sensor data as described above. Such services may become more common as more users decide to make use of the protection mechanisms available to them (i.e. making use of the ATT options on iOS, or re-setting their Ad ID on Android). Fingerprinting mobile devices may in some ways be harder than fingerprinting desktop devices, due to the fact that at the device level, mobile devices have less variety and are more standardised, but they are still ultimately eminently fingerprint-able \cite{hupperich2015robustness}.

\subsubsection{Third-party Mobile SDKs}

In much the same way that third-party tracking on the web relies on first-party developers embedding third-party code in their websites (or remotely loading it during a session), third-party tracking in mobile apps relies on developers embedding third-party code in the form of `libraries' and `Software Development Kits' (SDKs) \cite{book2013case}. Many of these SDKs are set up so that by default they connect to a third-party server and start sending personal data to it. This is made easier by the fact that permissions in Android are given on a per-app basis, rather than per-third-party tracker. As such, there is no way for users to grant permission to a first-party app developer but withhold permission for a third-party tracker embedded on that first-party app. Additionally, apps typically ratchet up the privacy-risking permissions they request over the lifetime of their installation, making additional requests on average every three months ~\cite{taylor2017}. This means that even if a user were to carefully consider whether they would grant requested permissions to an app before installing it, they cannot easily prevent onward use of such permissions by third-parties and are likely to be subjected to further sensitive requests after installation anyway.

Third-party trackers have taken advantage of such favourable conditions on mobile apps. Large-scale measurement studies of tracking on mobile apps have revealed a rich and varied third-party tracking ecosystem \cite{zang-2015-knows}. Many of these have been used to facilitate location-based ad targeting\cite{book2015empirical}. A study of the use of the Facebook SDK in Android apps, by Privacy International  \cite{privacy_international_how_2018}, found that 61\% of apps with the Facebook SDK automatically transfer data to Facebook the moment a user opens the app. This is regardless of whether they are logged into Facebook and even if they don't have a Facebook account. This means that Facebook knows immediately that a user has installed a particular app, even before any other data is transferred. Given the sensitivity of certain apps observed in this study - including a Muslim prayer app, a menstruation logging app, and apps aimed at young children - this alone could be quite revealing. The Android Ad ID was typically also automatically transferred, and many apps also send highly detailed and sensitive data to Facebook without asking the user. A more recent study of the same phenomenon found apps contact 4.7 trackers on average upon launching, without any user interaction \cite{kollnig2021fait}.

\subsubsection{Mobile Tracking Detection Methodology}
As with large-scale web privacy measurement, a variety of methods exist for detecting third-party trackers in apps. One approach, which is similar to web tracking measurement methods, involves `dynamic' network traffic monitoring. This involves setting up a real or virtual mobile OS, inspecting network traffic from the device and identifying any third-party destinations that relate to tracking. This can be achieved in different ways. One common approach has been OS-level instrumentation, with tools like TaintDroid~\cite{enck2014taintdroid}, and AppTrace~\cite{qiu2015apptrace}. An alternative is to analyse all communications traffic transmitted by an app whilst it is in use~\cite{ren2016recon}. In either case, apps must be used in order to trigger the tracking behaviour of interest. This can be acheived through simulated interactions via UI `monkeys' - bots programmed to interact with the app like a real user would \cite{jin2018they, binns_measuring_2018} - or through real user interactions with network traffic routed via a VPN \cite{shuba2016antmonitor,razaghpanah2018apps,song2015privacyguard}.

An alternative approach is to use `static' methods. These involve unpacking an application's source code (on Android systems, this comes as an Android application package(APK)) and detecting the use of third-party tracking within the source code~\cite{arzt2014flowdroid,batyuk2011using,egele2011pios,lin2014privacygrade}. Methods for detection vary, from simply searching for known tracker-associated hostnames to more complex machine learning approaches. The static methodologies can scale more easily as apps do not need to be actually run and interacted with in order to detect the use of third-party trackers. This means that analysis can be completed on an app within seconds rather than the minutes required for a real or simulated user session in the dynamic approach.

However, both methods are likely to over and under-state the presence of certain trackers. Dynamic approaches may miss third-party trackers that are embedded in the app if they fail to trigger the relevant user interactions or do not run for long enough. Static approaches may fail to identify trackers which are dynamically loaded from a remote server during run-time and therefore do not appear in the static code analysis. Furthermore, static approaches may struggle to detect the presence of third-party tracking libraries where the developer has obfuscated the app code. Some methods attempt to defeat such obfuscation attempts, e.g. using machine learning \cite{ma2016libradar}.

To understand the relative strengths and weaknesses of these two approaches, Binns et al. directly compared dynamic and static approaches on a set of 200 apps\cite{binns_measuring_2018}. They measured the extent of overlap between the two approaches, and how many trackers are left out by each approach. They found that on average, dynamic methods yielded 2.26 trackers which the static methods did not detect. Static methods yielded an average of 4.85 trackers which the dynamic methods did not detect. The average size of the intersection of mobile and web trackers found by each method was 2.9. While the static method had a higher positive rate, it is not possible to conclude that one method is better than the other without ground truth on the proportions of true positives detected by either method. Nevertheless, it gives some indication of the results that can be expected by either method.


\subsubsection{Distribution of App Trackers}
Both dynamic and static methods have been deployed to study the distribution of third-party tracking on mobile apps \cite{binns_measuring_2018,razaghpanah2018apps,vallina2016tracking}. A consistent finding is that, like the web, there are a few companies which have very large prevalence across first-party apps.  Alphabet / Google is consistently identified as the most prevalent and prominent on both platforms. Recent studies estimate that over 85\% of Android apps communicate with Google / Alphabet services \cite{krupp2021analysis,kollnig2021iphones}, with Facebook a distant but comfortable second. The dominance of these two trackers on Android has been well established and confirmed in multiple large scale studies of the Google Play Store, beginning in 2014 \cite{playdrone_2014}.

Large scale studies of the distribution of trackers tend to focus more on Android, in part because of the difficulties in studying iOS apps due to the locked-down nature of the iOS system. However, some such studies exist. A 2011 study applied static analysis methods to iOS apps, finding more than half collected unique device identifers and 55\% contained advertising or analytics libraries\cite{pios_2011}. A 2013 study of 226,000 iOS apps found that many apps access device identifiers, location and contacts \cite{agarwal_protectmyprivacy_2013}; subsequent iOS releases implemented permission requests to make such access conditional on user consent. Further studies have compared both mobile OS ecosystems against each other (e.g. \cite{han_comparing_2013,ren_recon_2016,kollnig2021iphones}). While Kollnig et al. found that the number of third party trackers present in iOS apps was somewhat lower than Android, neither platform was a clear `winner' on any of the facets of privacy studied \cite{kollnig2021iphones}.

\subsubsection{Comparing App Tracking Against Web Tracking}
One finding of large scale third-party tracking studies is that the distribution of trackers differs between the web and mobile. There are some trackers which are unique to mobile apps, and others which are unique to the web \cite{binns_measuring_2018,razaghpanah2018apps,vallina2016tracking}. This means that existing web-focused lists of trackers have good coverage of the tracking ecosystem on the web, but have low coverage for trackers in the mobile tracking ecosystem ~\cite{vallina2016tracking}. As with the web, mobile trackers are used for different purposes, from targeted advertising, to analytics, to security and more. Inferring the purposes of different trackers can be challenging, although it is often possible to infer their purpose from the way developers name functions and create URL paths, and metadata e.g. about the domain name. Several proposals aim to automatically infer such purposes from such information using machine learning \cite{jin2018they,shuba2020nomoats}.

Some studies have compared the extent of third-party tracking between web and mobile \cite{binns-2018-power,leung2016recon}. While neither platform is clearly better or worse than the other, there are meaningful differences between them. A comparison of 50 websites and their app equivalents found that while the website version of a service connected to more third-party domains, more device identifiers were leaked by apps \cite{leung2016recon}. For certain services, the app or the website offered a less privacy-invasive option, but there was no clear winner. A comparison of 300 services with both web and app versions found similar numbers of trackers between platforms, but quite different trackers were used\cite{binns_measuring_2018}.

\subsubsection{App Tracking and the Law}

Several studies have examined app tracking behaviours in light of regulatory requirements on app developers and third-parties. Reyes et al. found that a majority of the most popular free children's apps in the US are likely in violation of federal privacy law (Children's Online Privacy Protection Act 1998 (COPPA))\cite{reyes2018won}. The violation is due to third-party SDKs which, while capable of being configured in a COPPA-compliant way, are typically configured by children's app developers in ways which enable unlawful tracking and behavioural targeting of children. The researchers also found that 19\% of these apps use SDKs whose own terms of service outright prohibit their use on apps targeted at children.

In the context of EU data protection law, several studies have raised alarms around potential non-compliance by apps, trackers and platforms. Several studies suggests that many apps share data with trackers based in jurisdictions outside the EU, which are not deemed `adequate' by the European Commission in their protection of personal data and apparently without necessary legal protections in place \cite{binns-2018-mobile,razaghpanah2018apps}. Apps aimed at children are often those with the highest numbers of trackers\cite{binns-2018-mobile}, despite the European Data Protection Board advising that profiling of children for marketing purposes should be avoided.\footnote{`Because children represent a more vulnerable group of society, organisations should, in general, refrain from profiling them for marketing purposes' \cite{party2017article}} Finally, many apps which are required under GDPR and the ePrivacy Directive to seek consent before accessing data on a user's device, access such data without seeking consent and immediately share it with multiple third-party trackers (as found in studies mentioned above\cite{privacy_international_how_2018,kollnig2021fait}). Despite hopes that the introduction of the GDPR in 2018 would reduce the extent and distribution of third party trackers, a large-scale comparison of the Google Play Store before (2017) and after (2020) found very little change in their distribution on Android apps\cite{kollnig2021before}. 

\subsubsection{App Developer Incentives}

While end users have some power to change tracking practices, by refusing to use services with unwanted tracking, user pressure is ultimately unlikely to result in meaningful change given the rampant level of tracking across most first-parties and the practical necessity of using certain services to engage with essential services. As such, first-parties, and specifically the developers who make design choices about which third-party trackers to include, may have more influence over the extent of tracking in practice. Privacy researchers seeking an alternative point of leverage for addressing the negative externalities of third-party tracking have therefore sought to understand developer's reasons for integrating third-party trackers, their understanding of the privacy implications for their users, and the usability challenges they might face in adopting more privacy-preserving alternatives \cite{balebako_privacy_2014}. This is part of a broader effort to study developer perspectives as they are increasingly recognised as being key to better security and privacy \cite{acar2016you,assal_think_2019}. 

In the first major study of smartphone app developers in 2014, Balebako found that while third-party tracking SDKs were ubiquitous and heavily used, developers often were not even aware of the data collected by these tools \cite{balebako_privacy_2014}. More recent studies suggest that while developers are now more aware that by using third-party SDKs they are subjecting their users to potentially sensitive and invasive data flows, they see no other way to monetize their apps\cite{mhaidli_we_2019,ekambaranathan2021money}. They are resigned to the existing business models, and typically do not alter the tracker SDKs default, privacy-violating configurations. Even in the more sensitive context of children's apps, while developers do respect children's rights, they often feel they have no other viable business model than monetising user data through third-party SDKs \cite{ekambaranathan2021money}. Furthermore, the guidance provided by third-parties to developers is often lacking in detail and appears deliberately vague on how they are expected to implement the third-party tracking technology in legally compliant ways; one study of popular Android tracking SDKs found very limited and inadequate guidance on how developers were supposed to handle consent in compliance with EU data protection law\cite{kollnig2021fait}.

The changes to the iOS and (to a lesser extent) Google Play stores ecosystem (as described in section \ref{section:mobile_ad_id}), may ultimately have the biggest effect on developers' relationships with third party trackers. In addition to the ATT framework, under iOS version 14, developers must provide standardised privacy information which is displayed in the App store in the form of app privacy `labels' (similar to those proposed in HCI research, e.g. \cite{kelley2009nutrition}). Both measures may provide greater incentives for app developers to assess their relationships with third-parties, potentially cutting off those which provide little value to them, especially where they are worried about the effect of the new transparency measures and easy tracker-blocking options available to users.

\subsection{Internet-of-Things and Smart Homes}
\label{section:iot_counter}

Tracking and various other forms of surveillance existed before the web and mobile apps, and continue to operate outside them. However, the kind of tracking that was developed on the web and mobile, has provided a technical and commercial blueprint for tracking via devices outside those specific platforms. In recent years, a set of much-hyped and partially adopted technologies, grouped under the `Internet-of-Things' moniker (and in the domestic context, the `smart home'), have presented a novel frontier for the extension of third-party tracking. What makes an object `smart' is of course vague and nebulous \cite{kim2019definitions}, but one of the commonly cited criteria is that a smart device is internet-connected \cite{seymour2020strangers}. This ability to make network connections has allowed smart IoT devices to mimic or even embrace wholesale the same tracking infrastructures as already established in web and mobile-based tracking. Often, the very same companies providing web tracking technology are offering the same models for IoT \cite{weber2010internet}.

These different smart devices raise different privacy implications owing to their distinct locations, types of data collected, and modalities for interaction\cite{quinlan2019connected}. For instance, several automobile manufacturers including Volvo and Audi, have adopted Android Automobile, an automobile-focused version of Android, developed by Google. This uses the same app distribution model as the standard Android Play Store. This means that users must log in with their Google account, and developers creating apps for these vehicles are able to embed the full range of third-party trackers that are used in mobile apps, but have access to more data about another dimension of the user's life (i.e. their car journeys).

Various research has investigated the extent of tracking on IoT and smart devices, including smart toys~\cite{mcreynolds2017toys,chu2018security}, smart TVs~\cite{malkin2018can,mohajeri2019watching,varmarken2020tv}, robotic home assistants~\cite{urquhart2019responsible}, smart energy grid technology~\cite{goulden2014smart} and smart meter agents~\cite{costanza2014doing,rodden2013home}. As with web and mobile, a picture of the extent of third-party tracking can be gained from studying the network traffic from IoT devices. A 2019 study found that 57.45\% of the overall destinations contacted by US IoT devices are third or support parties (50.27\% for UK devices)\cite{ren2019information}.

The privacy implications of `smart' speakers and voice assistants like Amazon's Echo and Google Home, have also been the subject of multiple studies \cite{chung2017alexa,purington2017alexa}. Voice assistants placed in the home may be interacted with by multiple different users, providing insights not only to a single user's habits and interests but also those of their family, friends, and potentially any visitor to their home\cite{j2019exploring,chalhoub2021did}. Even if they only `listen' when addressed by their `wake word', they may end up responding accidentally when the wake word is said accidentally or the speaker makes a false positive error\cite{sun2020alexa}. While Amazon and Google at present deny using data from the Amazon Echo and Google Home smart speakers within their ad targeting services, third-party apps that are delivered via smart speakers, like the music streaming service Pandora, advertise the ability to target smart speaker users. Furthermore, existing adtech tracking companies advertise the ability to identify which consumers own smart speakers (which can be detected when their mobile devices pair with the smart speaker), so that they can be targeted for relevant services.\footnote{\url{https://perma.cc/24H9-J3JU}}

`Smart TV' platforms like Roku and Amazon Fire, so-called `over-the-top' services delivered via devices that plug into a TV or monitor, have their own `app store' ecosystems with channels or apps. These apps also have trackers embedded in them. Many of the companies behind them are already known from their presence in the web and mobile tracking ecosystems, but there are also some Smart TV-specific trackers. A study of the smart TV platform ecosystem and found hundreds of smart TV apps which exfiltrate personal data to third-parties and platform domains\cite{varmarken2020tv}. Another study found that 69\% of Roku channels and 89\% of
Amazon Fire TV channels sent traffic to known tracker domains \cite{mohajeri2019watching}. They also found that the platform-provided controls to limit tracking were practically ineffective, finding that having `limit ad tracking' setting on actually increased the number of tracker servers contacted.

Approaches to identifying tracking in IoT and smart devices are similar in some ways to those deployed on web and mobile devices; static and dynamic analyses are possible (see e.g. \cite{chu2018security} for a study of children's toys utilising both methods). However, each approach may require additional efforts in the IoT context. Static analysis requires access to the smart device's binary code. Unlike a website or app, the binary code is likely only accessible by physically prising open the device, locating and connecting to the flash chip and extracting the device firmware (for a systematic review of such techniques, see \cite{vasile2018breaking}). As a result, dynamic analysis may be preferable method for measurement of trackers. Some types of dynamic analysis used on web and mobile traffic may be equally applicable to IoT devices, where the analysis only observes network traffic source, destination, frequency, etc., but doesn't require decryption of the payloads. However, inspecting the \emph{content} of such traffic is difficult, because unlike a web browser or mobile device, IoT devices generally cannot be configured to trust a self-generated SSL certificate that would enable decryption of intercepted traffic.


The potential for tracker identification and blocking of tracker-related connections from IoT devices has been explored in multiple studies. Typically, these approaches are designed to be deployed at the router level, since most IoT devices, at least in a domestic setting, connect via the router. Various tools have been developed to provide analysis capability. Device fingerprinting tools like IoTSense~\cite{bezawada2018iotsense} and IoTSentinel~\cite{miettinen2017iot} allow for the reliable identification of devices based on their network traffic. Classifying behaviours of devices themselves is also possible with tools like HomeSnitch~\cite{oConnor2019homesnitch}, Peek-a-Boo~\cite{acar2018peek}, PingPong~\cite{trimananda2019pingpong}, and HoMonit~\cite{zhang2018homonit}. These can be used to infer \emph{why} a device is sending data to a particular destination at a particular time. When tracker-related traffic is detected, it can potentially be blocked at the router level, using systems like piHole\cite{sanoaf2018network}, a `Linux network-level advertisement and Internet tracker blocking application which acts as a DNS sinkhole and optionally a DHCP server, intended for use on a private network'. However these require some specialist technical expertise and so are effectively only an option for the rare few skilled and motivated users. An additional challenge here is blocking tracking without interfering with the useful and desired functionality of the device; Mandalari et al. propose IoTrimmer, a tool to automatically classify critical vs non-critical traffic and block the latter\cite{mandalari2020towards}, enabling the smart devices to continue working but without the tracking.

These router-level tools for tracker identification and control are typically research tools or targeted at users with higher technical skill. Some research has explored the potential for smart home privacy tools for non-expert users to better understand and exercise control over the flows of data from their devices. Seymour et al. developed a prototype privacy assistant, called Aretha, which combines a network disaggregator (to show smart device network flows), a personal tutor (to help users understand their significance), and a firewall (to selectively block unwanted traffic).\cite{seymour2020informing} Through a longitudinal study, where Aretha was installed in households over six weeks, they found that while most users did not engage with the firewall capability, when provided with the right kind of educational scaffolding,  people were able to formulate their own informed attitudes and strategies to address tracking in the smart home. Through interviews with users, Colnago et al. found that a balance needs to be struck between empowering users with controls, whilst not overwhelming them with information \cite{colnago2020informing}. Crabtree et al. argue that in order to make the IoT accountable to users, and compliant with the GDPR, data processing should be moved to the `edge' rather than being shared with first and third-parties, and propose the `Databox' as an edge device controlled by the user\cite{crabtree2018building}.




\subsection{Cross-device Tracking and `De-Anonymised' Identifiers}
\label{section:crossdevice}
Having covered how tracking works on the web, mobile apps, and other `smart' devices, we can now turn to how tracking operates across and between these different devices. Even if tracking on each platform works with slightly different mechanisms, these can be and are joined up behind the scenes \cite{zimmeck2017privacy}. For instance, fingerprinting is used for identifying a user as they move between an app and the mobile web, as a way to link together the cookie-based web tracking ecosystem with the Ad ID-based mobile tracking ecosystem \cite{das2016tracking}.

To paint a picture of how this all fits together, recall our example of Alice. Her web browsing activity from multiple websites was collated to form a profile, inferring that she may have an interest in ornithology, which could be used to target her via a web-based targeted advertisement. This profile, collated through cookies in her web browser, exists in a back-end database of a web-focused tracking service. Now, imagine Alice downloads an app on her Android device; as a result, her Android advertising ID is sent to multiple trackers, alongside various other personal data including her location, email address, and device details (handset model, OS version, network, etc.). One of those trackers has a commercial relationship with the web-focused tracking service, and the two parties seek to automatically match up their records. Their system finds a high-probability match based on Alice's location, email address and device details. Now Alice's cookie ID has been matched to her Android AdID and both trackers now know how to find Alice on either platform. This process is called `probabilistic' cross-device identification because the matching process is not guaranteed to be correct but rather has degrees of confidence depending on how many data points are available to narrow down Alice's identity. By contrast, `deterministic' cross-device identification involves an explicit identifier being passed from a service on one device or platform to another service. For instance, a web link in an app might allow the app to pass the Android advertising ID in the URL query string. When Alice clicks the link in the app, the URL \url{example.com/ad_ID?=12345} passes Alice's ad ID to \url{example.com}, where it can be associated with a cookie. Similar probabilistic and deterministic mechanisms may allow cross-device tracking between Alice's smart home devices and her web and mobile presence.

A large data broker industry exists for the purposes of linking mobile device identifiers, real names, physical addresses, phone numbers, email addresses, and IP addresses together.\footnote{https://perma.cc/F8S8-5FNY} While many maintain the pretence that mobile advertising identifiers are `anonymous', the widespread practices of the adtech sector ensure that they are in practice highly identifiable to a wide variety of actors. Such identifiers are routinely broadcast to hundreds of companies participating in the programmatic adtech real-time-bidding ecosystem as soon as an app is installed and run. Additional so-called `anonymisation' measures are applied by many tracking companies, which consist of converting personal data into a supposedly less easily identifiable pseudonymous form. This is often acheived by running the personal data through a hash function, which maps input data of an arbitrary size to a seemingly random fixed output (e.g. under the SHA-256 hash algorithm, `user123@example.com' becomes `865c4dd20427a134c6a6637f1d2c905077b7d7380cd8e029c4a358e222f91b43'). However, anyone with knowledge of the hashing method can check if a given identifier results in a matching hash. So if a tracker shares a hash of `user123' with a data broker, then the broker (or anyone else they share it with) can attempt to match the hash against the hashes of all the user identifiers they already have. Any matches then allow further enrichment of a user profile associated with that hash. Effectively, a hash of an identifier can become every bit as effective as the identifier itself, acting as a key to match up lots of different information about a person from various sources, all under the pretence of `anonymity'. Standard industry approaches to hashing identifiers make such matching processes trivially easy \cite{narayanan2016precautionary}.

Many companies may be participating in real-time-bidding not because they actually want to bid on the available auctions, but simply for the opportunity to receive personal data through the bid stream and monetise it in various ways. Examples of how such data can be easily bought abound.\footnote{\url{https://perma.cc/FUY9-3UGG}} In July 2021, Catholic news outlet The Pillar published a report detailing the private romantic life of a high-ranking Roman Catholic official, who subsequently resigned. The report was based on data purchased from a data broker, which had originally been obtained via a third-party tracker present on the gay dating app Grindr. This dataset featured personal data of Grindr users, including GPS location data, which enabled the official to be identified.\footnote{https://perma.cc/3A67-K5SE} This is just one of a large number of companies offering such de-anonymization services, who use terms like `identity resolution' to describe their services. Ultimately, the disparate tracking infrastructures embedded in platforms can often be tied together to form individual profiles of identifiable individuals, by firms who sell their services to a range of customers with seemingly little scrutiny. These profiles are likely to contain all kinds of personal data: an individual's purchase history; medical concerns and hospital visits; apps installed; media subscriptions and loyalty programs; places of work and socialising; political views and religious beliefs; debts and loan status; tastes in food and diet; interactions with call centres; bills and invoices; survey responses; attitude and lifestyle; and much else besides.




\section{Whither Tracking?}

Previous sections covered the history of tracking and some major strands of research from computer science and related disciplines. This section takes a more policy-oriented and speculative perspective, briefly introducing and discussing some of the contemporary issues and trends --- economic, legal, and technical --- which appear likely to influence the long term trajectory of tracking.

\subsection{An Adtech Market Crash?}

One factor which may have significant effects on the future of tracking is the possibility of a major downturn in the advertising technology industry. Some of the factors that might potentially drive this outcome have already been mentioned in passing above, including: widespread opposition by end-users; increasing use of tracker blocking tools and tracker-hostile measures by browsers and smartphone OS providers; efforts by developers to ditch privacy-invasive third-party tools; and the adoption of privacy-preserving alternatives to advertising and analytics, which potentially remove the demand for intermediaries in their current form. Several other factors can be added to these.

First, while the adtech industry is now estimated to be worth around \$455 billion, it is often unclear how much value the intermediaries in the middle provide to advertisers and whether they pass a fair cut on to the first-party publishers whose audiences' attention they monetise. It is particularly hard to measure the returns on digital advertising with any confidence; a study involving 25 field experiments was only able to estimate returns with a 100 percentage-points-wide confidence interval \cite{lewis-2014-unfavorable}. It is also unclear how much value is added by tracking individual users, compared to targeting adverts based on the context (e.g. targeting adverts based on the content of the page rather than information about the user visiting it). A 2019 study found that use of tracking cookie data to target ads only yielded a 4\% increase in revenue for publishers compared to contextual targeting alone \cite{marotta2019online}, and results of limited experiments from publishers are also promising. \footnote{\url{https://perma.cc/XNK9-F9B9}.}

Furthermore, first-parties who sell ad space appear to receive a surprisingly small portion of the amount spent by advertisers, with the rest going to various adtech intermediaries. A report from the Incorporated Society of British Advertisers (ISBA) estimated that for every dollar spent on adtech, about half (51 cents) goes to the first-party website who displays the ad, a third is taken by the adtech intermediaries, and a final average of 15 cents was impossible to account for; as Gizmodo reporter Shoshana Wodinsky described, this is `a Bermuda triangle at the center of the web where these billions of dollars just... vanished.'\footnote{\url{https://perma.cc/2G39-MZ7Q}}

Another factor is the widespread and extensive levels of fraud \cite{watkins2019guide}. Digital advertising fraud typically involves a fraudster creating fake ad traffic using automated robots or `bots' which mimic the interactions of human users. These generate ad impressions which inflate the advertising campaign metrics, despite no human ever seeing the ad itself. Bot-operators, in cahoots with first or third-parties, then charge advertisers for these wasted ad impressions. According to one recent estimate, ad fraud causes economic losses of 35 billion US dollars per year.\footnote{\url{https://perma.cc/PNC9-BDGU}}. Some research exists demonstrating ad fraud in Android applications \cite{crussell2014madfraud,liu2014decaf}, but the full extent is uncertain.

The lack of clarity about the value actually provided within the adtech industry, and the apparently high levels of ad fraud, has lead some to question whether adtech vendors may soon lose the confidence of their customers and investors. Drawing parallels to the 2008 financial crash, Tim Hwang has argued that this could trigger a `sub-prime attention crisis'\cite{hwang2020subprime}. While some may relish the prospect of a collapse of adtech, the fallout might precipitate more worrying developments. Investors in tracking companies will demand a return on their investment; if adtech revenue collapses, these companies will look to other ways to turn a profit from the large amounts of highly sensitive data they have amassed, which might be cause even greater harm. To avoid such outcomes, the collapse of adtech revenues would need to go hand in hand with a dismantling of their underlying surveillance infrastructure.

Furthermore, while problematic, adtech currently provides revenue for an industry with an important societal function: the news media \cite{libert-2015-think,libert2019good}. An adtech crash could therefore undermine the `fourth estate'. That said, one could argue that the function of the press as a check on power is already undermined by widespread tracking \cite{christians-2009-normative,bennett-2015-introduction}, because readers need to be able to read material critical of the state or other powerful actors  without the fear of being watched by them \cite{stoycheff-2016-under,marthews-2015-government,america-2013-chilling,penney-2016-chilling, pew-2014-privacy_post_snowden} (what Neil Richards calls `intellectual privacy' \cite{richards-2008-intellectual}). Alternative sources of funding for news media may therefore be important to establish whether or not adtech survives in its current form. Some alternatives include subscriptions, more consensual and privacy-respecting targeting within the first-party context, and `contextual' advertising as explained above.

\subsection{Data protection and privacy law}

Laws around the collection and use of personal data emerged in many countries in the 1970's \cite{fuster2014emergence}. These were based in part on the Fair Information Practice Principles (FIPPS) developed in a 1973 report by the US Department of Health, Education and Welfare \cite{gellman-2016-fair,ware1973records}. Various European countries enacted proto-data protection instruments, starting with the Swedish Data Act in 1973.
In the EU, data protection law aims to protect fundamental rights and freedoms in relation to the processing of personal data.


Some firms engaged in the tracking industry have claimed that data protection law does not apply to them because data they process is not `personal data'; instead, they claim, it is `anonymous' and cannot be used to identify an individual. However, such claims are based on an erroneously narrow understanding of personal data as `personally identifying information' (a term used in US law), typically meaning a legal name, email address, social security number, etc. This is much narrower than the definition of personal data in EU and other jurisdictions, which includes any information that could be used, alone or in combination with other data, to identify an individual or single them out. On this broader definition, a large portion of the data collected by third-party trackers is personal data, either because it is a unique identifier, or because it alone or combined with other data suffices to single out an individual. So a cookie with a user's unique identifier which serves to distinguish them from other users is personal data, even if it doesn't include their name or email address. But also, a set of browser configuration settings or series of location co-ordinates, if they enable data controller to single that user out, could be personal data\cite{borgesius2016singling}. Furthermore, so-called `anonymisation' as practiced by many tracking companies (as described in section \ref{section:crossdevice}) is actually only `pseudonymisation' under data protection law; converting personal data into a less easily identifiable, but still identifiable, pseudonymous form. As a result, much of this data would be personal data under the GDPR.

E.U. case law has clarified that what counts as `personal data' is contextual; an Internet Protocol (IP) address may be personal data if it is typically associated with an individual (e.g. if it is associated with their home internet connection), but may not if used by multiple different people (e.g. for a public WiFi hotspot) \cite{borgesius2017breyer}. Based on analysis of such case law and the guidance issued by data protection authorities, Purtova has argued that `identification' in data protection law extends also to instances of `personalisation'; where data is used to personalise a service to someone by relatively uniquely characterising them by mapping them `in relation to multiple dimensions within a multidimensional space'\cite{purtova2021knowing}. Whether this expansive interpretation of the scope of personal data holds up in future judgements remains to be seen, but even a less expansive interpretation would arguably still capture the vast majority of tracking practices described above.

Another area of controversy in the application of data protection to tracking is around the concepts of controller and processor. First-party services are typically data controllers, because they decide the purposes of processing personal data of their users. Third-party trackers may be either controllers or processors, depending on whether they operate under the instruction of first-parties. For instance, a third-party analytics provider might process personal data on behalf of the first-party according to the first-party's instructions, and thus be a processor. Alternatively, were that third-party analytics provider to use the personal data from the first-party to create additional services outside of their instructions from the first-party, e.g. to develop services or improve their AI models, they would likely become a controller, with a greater level of responsibility.

Many third-parties therefore attempt to minimise their compliance burden by claiming to be mere processors; by presenting take-it-or-leave-it contractual terms, according to which the first-party agrees to `instruct' the third-party to undertake the processing the third-party wanted to do anyway, but without taking on responsibility as controller \cite{cobbe2021artificial}. The resulting situation is that powerful third-party trackers, who in reality decide on the purposes and means of processing and therefore ought to be classed as controllers, claim to be mere processors acting on instruction of first-parties; a kind of `puppet controllership' where they exercise power without the regulatory compliance burden, which is pushed onto the first-party. However, recent case law of the Court of Justice of the European Union affirms that the bar may indeed be low enough to qualify many third-parties as controllers (or joint controllers where they jointly decide on the purposes and means of processing).\footnote{See e.g. \emph{Case C-49/17,Fashion ID, 2019 ECLI:EU:C:2019:629}, finding that when a website embeds a Facebook `Like' button, which facilitates third-party tracking, it is a joint controller with Facebook; and \emph{Case C-210/16, Unabhängiges Landeszentrum für Datenschutz Schleswig-Holstein v. Wirtschaftsakademie Schleswig-Holstein GmbH}, where the operator of a Facebook Fanpage operator was deemed a joint controller.} Relatedly, third party consent management platforms are also likely to be considered controllers under data protection law \cite{santos2021consent}, which could lead to difficulties regarding legal liability in the context of non-compliant tracking facilitated by first and third parties.

Finally, the real-time bidding architecture deployed by the adtech industry, discussed in section 2.3 is argued to be in violation of the transparency and security requirements of data protection law.\footnote{See e.g. \url{https://perma.cc/YQ55-EF33}} These arguments focus on three main ways in which real-time bidding (RTB) violates data protection law, all arising from the way that personal data is routinely broadcast in the process of an RTB auction to hundreds of competing bidders \cite{veale2021adtech}. First, intermediaries in the RTB ecosystem typically require consent but are unable to gain it validly because data subjects cannot meaningfully consent to the processing of their data by thousands of potential bidders; second, even if they could use some lawful basis other than consent, they would fail to meet the transparency requirements that would still apply, for the same reason; third, it is impractical to vet each recipient and ensure the security of such data, but failure to do so arguably violates the security principle of the GDPR.

Compared to EU data protection law, US privacy law is a patchwork of federal and sectoral laws, and state-level instruments. Existing federal privacy laws include the Health Insurance Portability and Accountability Act of 1996, Children’s Online Privacy Protection Act of 1998, and 1999 Gramm-Leach Bliley Act. However, for many years, the web advertising technology industry has intensively lobbied against regulation and in favour of self-regulation \cite{iab-2010-self_reg, ftc-2009-oba}, and largely escaped the former at least during the 2000's \cite{culnan-2000-protecting} and early 2010's. This approach was largely supported by regulators who endorsed self-regulatory principles \cite{ftc-2009-oba}. Industry-led intiatives were proposed to limit the effects of tracking, such as the `ad choices' button attached to online advertising \cite{komanduri-2011-adchoices}, which allowed users to decide what ads they wanted to see (but not to actually limit the underlying third-party tracking). This exclusively self-regulatory approach was widely regarded as a failure by privacy advocates \cite{cate-2006-consumer}.

Federal privacy law applies to government use of data, and web profiling is a private enterprise. However, national security agencies utilise this private infrastructure, either through warrants or hacking \cite{soltani-2013-nsa_google}, while law enforcement are known to purchase data from data brokers \cite{hoofnagle-2003-big}. Private sector use of data has been addressed in some cases by the Federal Trade Commission (FTC), under their remit to regulate unfair and deceptive practices. For instance, in the context of tracking, practices which trick users into giving up their data or buying services by mistake \cite{mathur2018characterizing}, might be addressed within this doctrine. In 2013, the FTC investigated the data broker industry, driven by media reports of practices including the sale of lists of people with specific health conditions, victims of particular crimes, and more \cite{ussenate-2013-databrokers,wsj-2013-rapelist}.

In recent years, states have passed more comprehensive data privacy laws which, while more limited than the data protection regimes of the EU and many other countries, do contain meaningful constraints on personal data processing. For instance, the California Consumer Privacy Act (CCPA) which became effective on January 1st 2020, contains a set of individual rights and obligations on firms which mirror some of those found in the GDPR. It is regarded as `the strictest data privacy law in the US'.\footnote{\url{https://perma.cc/UZ7Y-EHLZ}} In the wake of such state-level laws, there is momentum for a federal consumer privacy law which would harmonise laws between states. It remains to be seen whether such a federal law will be passed, and if so, whether it would constitute a levelling-up to the high standards of the CCPA, or a lowering of the bar across the US.

In 2021, China adopted a new comprehensive data protection law, the `Personal Information Protection Law' \cite{determann2021china}, which could have significant consequences for online tracking, both for domestic Chinese companies, and for foreign companies operating in China. Users have the legal right to block algorithmically-curated information and personalised ads. Companies have to get affirmative user consent to serve ads based on personal information that has a `major influence on individual rights and interests' (Article 27). Enforcement under the new law has already been taken against various apps for privacy violations; after a warning the previous month, in December 106 apps were removed from a variety of app stores.\footnote{\url{https://perma.cc/XL8L-HZJU}}

\subsection{Competition and Antitrust}

Large tech firms have been the focus of recent attention by antitrust and competition regulators, in part due to their control over key parts of the tracking infrastructure. The ability to track people across the web and other platforms is concentrated in a few powerful players, especially Google with its Chrome Browser and Android OS. Competition regulators in multiple jurisdictions have responded by launching investigations into mergers and acquisitions, considering the effects of consolidation of tracking capability on the market power of firms.

In some cases, these have been precipitated by proposals by large tech firms which purportedly aim to improve privacy. Specifically, Google's stated intention to phase out the use of third party cookies (following the lead of other major web browsers) has raised concerns about the potentially anti-competitive effects on rival adtech vendors who rely on third party cookies to facilitate tracking and targeting of advertising. In response, the EU Commission, UK Competition and Markets Authority (CMA), and Australian Competition and Consumer Commission, among others, initiated investigations into the move \cite{ft2021ukgerman}. That the proposed move away from third party cookies was ostensibly motivated by privacy concerns has caused some to pit data protection and competition policy against each other. However, the apparent conflict between data protection and competition needs to be set against a broader view of their fundamental aims. Arguably, both data protection and competition policy are concerned with constraining the power of corporations, and ensuring the public get the benefits of technology. Within this evolving policy debate, many now recognise that these two areas of regulation are more mutually supportive than they are conflicting\cite{lynskey2017aligning,lynskey2017aligning,buttarelli2019not}.

Still, some of the traditional approaches, concepts and tools of each regulatory domain may need revising in order to realise this mutually supportive relationship. One issue is the way in which competition law has traditionally measured market power, which is used to determine whether and how to intervene with respect to certain market actors. Traditional market share calculations will likely fail to capture the kind of power exercised by firms engaging in the collection of personal data; as the European Data Protection Supervisor noted in 2014, power over personal data ``cannot easily be calculated by reference to data on traditional sales or volume''~\cite{hustinx2014privacy}. Alternative market concentration measures and merger and acquisition review processes may be needed which take account of privacy as an aspect of quality~\cite{swire2007protecting}; and measure the combinations of data and computational infrastructure that could arise from the consolidated entity and what effects that could have on the structure of the market and potential harms to individuals\cite{binns_dissolving_2020}. In recent years regulators have begun to consider potential privacy harms arising from the consolidation of tracking capabilities and the effects on privacy within in competition and antitrust activity ~\cite{schechner2017germany}.

The possibility for abuse of dominance and vertical integration is palpable in the case of a firm like Google / Alphabet. Not only does it run the most prominent third-party tracker network on Android, it also controls the app marketplace, the operating system and its standard applications. Meanwhile, on the web it also controls the most popular browser, search engine and third-party advertising and analytics services. These activities cannot be considered in isolation ~\cite{edelman2016android}; dominance in tracking might easily spill over into other spheres of business, and vice-versa \cite{edelman2015does}. Viewing large dominant firms with the power to shape markets as an inherent problem, regardless of their immediate impact on prices, is characteristic of a new wave of US antitrust academics and policymakers known as the `Neo-Brandeisians' (a revival of an earlier era of antitrust developed by Justice Brandeis) \cite{khan_amazons_2017,wu_curse_2018,pasquale2013privacy}.

The ultimate outcome of competition and antitrust regulation is uncertain. One possibility is structural separation of dominant tech firms, including separation of their tracking infrastructure from other parts of the business. For instance, Google might be forced to divide its third-party advertising and analytics networks from each other and its wider services. Other remedies might involve the forced sharing of data and / or infrastructure with competitors via interoperability measures defined by open protocols \cite{brown2020interoperability}. Where dominant firms have services which are ostensibly `on the side of the user', such as web browsers and mobile operating systems, specific statutory regulation and contractual measures could be created to address conflicts of interest arising from their other service offerings. Some have even suggested such `user agents' should be bound by fiduciary duties, providing structural leverage to correct the balance of power between individuals and tech companies\cite{berjon2021fiduciary}.

\subsection{`Privacy-Preserving' Tracking?}



In 2020, Google announced a set of proposals for a post-cookie, purpotedly privacy-respecting future for the web called the `Privacy Sandbox'.\footnote{See \url{https://perma.cc/88LD-PMA6}} There are various proposals within the Sandbox, but one in particular raises important technical, legal and conceptual questions about the possible future of tracking. Called `Federated Learning of Cohorts' (FLoC)\cite{yao2021federated,langheinrich2021floc}, it aims to replace current arrangements with third-party cookies, but still allow insights about a user to be inferred based on their membership in a group of similar users.

As covered in section \ref{section:pp_alternatives}, various alternatives to tracking have already been proposed which aim to enable targeting of content and ads, without the user's data leaving their browser. In such proposals, the browser application itself creates and stores a profile of the user's interests, and selects from a range of selected ads, revealing nothing to first and third parties about the user. Some browser vendors are exploring or beginning to implement such systems; for instance, the Brave browser is implementing a privacy-preserving decentralised ad platform called THEMIS \cite{pestana2020themis}. FLoC would be something of a halfway house between entirely local and private targeting, and the `broadcast' model currently used by real-time bidding adtech systems. It would use an in-browser classifier to detect, based on web page content, the categories of websites that a browser visits. These classifications would then be used to group web browsers together based on how similar the categories of websites they’ve visited are. When a user visits a website, instead of allowing third-party cookies being set and read, the user's FLoC cohort would be shared with third-parties. This would not allow them to uniquely identify the user (at least not on its own), but would allow them to target advertisements and tailor content to the user based on their membership in a FLoC cohort.



Such targeting would require first and third-party services to derive insights about those FLoCs, e.g. their demographics and what products they might be interested in (using auxiliary data, analytics, and machine learning). 
Through such processes, FLoC groups would become associated with categories of interest to advertisers, as well as potentially other commercial or state actors; e.g. health conditions, political opinions, or sexual orientation. This system might technically preserve privacy in the narrow sense that you can't be distinguished from others in your FLoC cohort; but the information, opportunities, and influences you are exposed to would be altered and shaped according to what can be inferred about you based on your FLoC cohort.

Trials of FLoC began with a select group of Chrome browser users in Spring 2021, but the initial response from privacy groups, researchers, and competing browser makers has been largely hostile. At the time of writing, it appears that other alternatives are being considered. A simpler system where instead of broadcasting a single numerical cohort ID (which adtech intermediaries and advertisers would have to infer the meaning of for themselves), the browser broadcasts topic IDs (e.g. `performing arts' or `fitness'), which would provide a simpler approach for targeting.\footnote{\url{https://perma.cc/4K7A-EKRC}} It remains to be seen whether this move is acceptable from a user perspective; while many might appreciate no longer having individual identifiers spread around the real-time bidding ecosystem, they might still be concerned that their algorithmically-derived interests are routinely broadcast to third parties. Furthermore, it is unlikely that FLoC would wholly replace the existing targeting systems; it is likely that both the old and new systems would operate side-by-side. In which case, \emph{both} individual identifiers \emph{and} FLoC cohorts / interest categories would be broadcast to websites and into the real-time bidding process; rather than improving privacy, FLoC could merely add additional data points to existing tracking processes. Other mechanisms would therefore still likely be needed to provide effective protection from fingerprint-based tracking.

\subsection{Concluding Remarks}

This article has aimed to provide an overview of a widespread and complex phenomenon, which --- despite decades of research, increasing coverage in the media, and attention by regulators --- remains under-examined given its societal importance. Many areas have not been touched on here in detail, from how tracking operates outside of the commonly-studied contexts of North America, Europe, and China, to the political economy of tracking infrastructure\cite{helles2020infrastructures}, and the use of tracking in sensitive areas like health\cite{libert-2015-bmj} and credit\cite{deville2019digital}. There are many differing and equally fascinating ways in which tracking on the web, mobile and IoT intersect with modern economic, social and political realities which this overview could not include. However, it hopefully provides sufficient background for a reader aiming to pursue more detailed research.

For all the technical complexity of the third-party tracking ecosystem, the countermeasures deployed by browsers and smartphone platforms, and the exotic new `privacy-preserving' alternatives, the underlying problems of tracking are political \cite{hildebrandt2008profiling,bayamiloglu2018being,gandy1993panoptic}. Who (if anyone), should have the power to target, to segment, to sort and persuade populations through the shaping of their online environments at scale? How should such power be constrained, re-distributed, or abolished? How might the infrastructure, resources, and expertise currently serving the tracking industry be diverted or transformed to serve alternative needs and ends?

\section{Acknowledgements}
Thanks to Kieron O'Hara for helping to formulate the scope and guide the shape of this article from its early stages, and subsequent feedback. Thanks to Konrad Kollnig for feedback on an early draft. Thanks to two anonymous reviewers for their valuable feedback, which substantially improved the paper.

\bibliographystyle{unsrt}  
\bibliography{references} 
\end{document}